\documentclass{emulateapj}
\usepackage{datetime}
\usepackage{amsmath}

\usepackage{color}

\newcommand{\eg}{{\it e.g.}}

\newcommand{\ie}{{\it i.e.}}

\newcommand{\teff}{$T_{\rm eff}$}

\newcommand{\ms}{{m~$\rm s^{-1}$}}

\newcommand{\mearth}{{$M_\oplus$}}
\newcommand{\rearth}{{$R_\oplus$}}
\newcommand{\rsun}{{$R_\odot$}}
\newcommand{\msun}{{$M_\odot$}}
\newcommand{\kepler}{{\it Kepler}}
\newcommand{\icarus}{{Icarus}}

\newcommand{\msini}{$M_p\sin{i}$}

\newcommand{\feh}{{[Fe/H]}~}

\begin{document}

\title{Miniature Exoplanet Radial Velocity Array (MINERVA) I. \\ Design, Commissioning, and First Science Results} 
\author{Jonathan J. Swift,\altaffilmark{1}, 
Michael Bottom\altaffilmark{1}, 
John A. Johnson\altaffilmark{2}, 
Jason T. Wright\altaffilmark{3}, 
Nate McCrady\altaffilmark{4}, 
Robert A. Wittenmyer\altaffilmark{5}, 
Peter Plavchan\altaffilmark{6},  
Reed Riddle\altaffilmark{1},  
Philip S. Muirhead\altaffilmark{7}, 
Erich Herzig\altaffilmark{1}, 
Justin Myles\altaffilmark{8}, 
Cullen H. Blake\altaffilmark{9}, 
Jason Eastman\altaffilmark{2}, 
Thomas G. Beatty\altaffilmark{3},    
Stuart I. Barnes\altaffilmark{10},
Steven R. Gibson\altaffilmark{11},
Brian Lin\altaffilmark{1}, 
Ming Zhao\altaffilmark{3},  
Paul Gardner\altaffilmark{1}, 
Emilio Falco\altaffilmark{12}, 
Stephen Criswell\altaffilmark{12}, 
Chantanelle Nava\altaffilmark{4}, 
Connor Robinson\altaffilmark{4}, 
David H. Sliski\altaffilmark{9}, 
Richard Hedrick\altaffilmark{13}, 
Kevin Ivarsen\altaffilmark{13}, 
Annie Hjelstrom\altaffilmark{14}, 
Jon de Vera\altaffilmark{14}, 
Andrew Szentgyorgyi\altaffilmark{12}}

\altaffiltext{1}{California Institute of Technology, 1200 E. California Blvd., Pasadena, CA  91125 USA}
\altaffiltext{2}{Harvard-Smithsonian Center for Astrophysics, Cambridge, MA 02138 USA}
\altaffiltext{3}{Department of Astronomy and Astrophysics and Center for Exoplanets and Habitable Worlds, The Pennsylvania State University, University Park, PA 16802}
\altaffiltext{4}{Department of Physics and Astronomy, University of Montana, 32 Campus Drive, No. 1080, Missoula, MT 59812 USA}
\altaffiltext{5}{School of Physics and Australian Centre for Astrobiology, UNSW Australia, Sydney, NSW 2052, Australia.}
\altaffiltext{6}{Department of Physics Astronomy and Materials Science, Missouri State University, 901 S. National Ave., Springfield, MO 65897 USA}
\altaffiltext{7}{Department of Astronomy, Boston University, 725 Commonwealth Ave., Boston, MA 02215 USA}
\altaffiltext{8}{Department of Astronomy, Yale University, New Haven, CT 06511 USA}
\altaffiltext{9}{The University of Pennsylvania, Department of Physics and Astronomy, Philadelphia, PA, 19104, USA}
\altaffiltext{10}{Stuart Barnes Optical Design, 1094NK Amsterdam, The Netherlands (current address: Leibniz-Institut f\"ur Astrophysik (AIP), 14482 Potsdam, Germany)}
\altaffiltext{11}{Department of Physics and Astronomy, University of Canterbury, Private Bag 4800, Christchurch 8140, New Zealand (current address: Space Sciences Laboratory, University of California, 7 Gauss Way, Berkeley, CA 94720)}
\altaffiltext{12}{Smithsonian Astrophysical Observatory, 60 Garden St, Cambridge, MA 02138 USA}
\altaffiltext{13}{PlaneWave Instruments Inc., 1819 Kona Drive, Rancho Dominguez, CA 90220 USA}
\altaffiltext{14}{Las Cumbres Observatory Global Telescope Network, 6740 Cortona Dr. Suite 102, Goleta, CA 93117 USA}

%


\begin{abstract}
The MINiature Exoplanet Radial Velocity Array (MINERVA) is a US-based observational facility dedicated to the discovery and characterization of exoplanets around a nearby sample of bright stars. MINERVA employs a robotic array of four 0.7\,m telescopes outfitted for both high-resolution spectroscopy and photometry, and is designed for completely autonomous operation. The primary science program is a dedicated radial velocity survey and the secondary science objective is to obtain high precision transit light curves. The modular design of the facility and the flexibility of our hardware allows for both science programs to be pursued simultaneously, while the robotic control software provides a robust and efficient means to carry out nightly observations. In this article, we describe the design of MINERVA including major hardware components, software, and science goals. The telescopes and photometry cameras are characterized at our test facility on the Caltech campus in Pasadena, CA, and their on-sky performance is validated. New observations from our test facility demonstrate sub-mmag photometric precision of one of our radial velocity survey targets, and we present new transit observations and fits of WASP-52b---a known hot-Jupiter with an inflated radius and misaligned orbit. The process of relocating the MINERVA hardware to its final destination at the Fred Lawrence Whipple Observatory in southern Arizona has begun, and science operations are expected to commence within 2015.
\end{abstract}


\maketitle 


\section{Motivation}
\label{sec:intro}
The field of exoplanetary science has experienced rapid growth since the discoveries of the first planetary--mass companions 
more than 2 decades ago \citep{Campbell1988,Latham1989,Wolszczan1992,Wolszczan1994,Mayor1995,Marcy1996}.
These initial discoveries  spawned myriad observational efforts that have expanded our view of planetary systems from a single example---our own Solar System---to a diverse statistical ensemble containing hundreds of confirmed systems and thousands of candidates \citep{eod2011,eod,Akeson2013}\footnote{also see \url{exoplanets.eu}}. Modern techniques for discovering and characterizing exoplanets include 
transits \citep{Rosenblatt1971,Borucki1984,Borucki1985,Henry2000, Charbonneau2000},  microlensing \citep{Mao1991,Gould1992,Bond04} and direct detection \citep{Marois2008,Kalas2008}. Our understanding of planet formation and evolution, and the possibility for other intelligent life in the cosmos has been transformed by this swift progress that continues to accelerate in exciting directions.

At the time of the launch of the \kepler\ Mission in 2009 \citep{Borucki2010}, radial velocity (RV) surveys had discovered more than 400 planets orbiting nearby stars by detecting the minute, periodic Doppler shifts in stellar spectra induced by orbiting planetary companions. The diverse collection of RV--detected planets revealed many important correlations between planet occurrence and stellar properties \citep{Santos2004,Fischer2005,Johnson2010}, as well as detailed information about the physical and orbital characteristics of planets outside our Solar System
(for a few interesting examples see references \citenum{Ford05,Correia09,Rivera2010}).
We now know that planets throughout the Galaxy have a wider range of masses, radii, and internal structures than the planets of our own Solar System. Of particular interest is a new class of planet---the so-called ``super-Earths''---with masses and radii intermediate to the Solar System terrestrial planets and the ice-giants Neptune and Uranus (see refs. \citenum{Santos2004,rivera05,valencia06,Valencia2007,Udry2007} for some early examples).

From the standpoint of searching for low-mass exoplanets that resemble those of the inner Solar System, one of the most exciting statistical results from RV-detected planets is a planetary mass function that rises steeply toward Earth-mass planets. Howard {\it et al.} (2010)\citep{Howard2010b} analyzed the planet discoveries and detection efficiency of the {\em NASA/UC $\eta_\oplus$ Survey} conducted at Keck Observatory using the HIRES spectrograph, and found that the number of planets per interval in $M_{\rm min} \equiv M_p\sin{i}$ scales as $dN/d\log{M_{\rm min}} \sim M_{\rm min}^{-0.48}$. Extrapolation of this relationship to terrestrial minimum masses ($0.5 < M_p\sin{i}/M_\oplus < 2.0$) and periods $P < 50$~days led to the remarkable prediction that 23\% of Sun-like stars harbor an Earth-mass planet. This number  agrees well with results from the HARPS surveys of chromospherically--quiet FGK dwarfs \citep{Lovis2009,mayor11} and M dwarfs \citep{bonfils13}. Small planets are far more common than large planets throughout the Galaxy, which bodes well for the search for Earth--like planets in the Solar Neighborhood. 

In the past four years, the more than 4000 planet candidates\footnote{as of 2014 September 1 on \url{http://exoplanetarchive.ipac.caltech.edu}} detected by the \kepler\ Space Mission \citep{Borucki2011a, Borucki2011b, Batalha2013, Burke2014} have strengthened the scientific results from RV surveys and greatly expanded our knowledge of exoplanet properties down to sizes comparable to, and even below the Earth \citep{Muirhead2012a, Barclay2013}. Consistent with the results of RV surveys, the \kepler\ discoveries strikingly illustrate that the number of planets increases rapidly with decreasing planet radius \citep{Howard2010a,Cassan2012,Fressin2013}. There is still uncertainty about how much this trend softens or even turns over toward the smallest detected planets \citep{Morton2014,Foreman-Mackey2014}. However, it is clear that there are more planets smaller than 4\,\rearth\ than larger ones in the Galaxy. Estimates of the occurrence of Earth-size planets around Sun-like stars range from 10 to 15\% \citep{Fressin2013,Petigura2013b} and the occurrence of Earth--size planets in Earth--like orbits is estimated to be between 2 and 6\% \citep{Foreman-Mackey2014,Petigura2013b}.

Less than a decade ago the only known terrestrial planets orbiting Main Sequence stars resided in our Solar System, and there was little expectation that, even if they existed around other stars, they would be presently discovered. Now \kepler\ has shown us that they are very common and may even constitute the dominant population of exoplanets, particularly if one considers recent planet occurrence estimates around low-mass, M-type dwarfs, which are the most numerous stars in the Galaxy \citep{Swift2013,Dressing2013,Morton2014}. While the the characteristic distance of a \kepler\ target is $\sim 1$\,kpc, the statistical results from the \kepler\ Mission should extend to the Solar Neighborhood, thereby informing us about the closest stars to Earth (but see ref.~\citenum{Wright12}). It is now clear that the night sky is teeming with unseen terrestrial-mass planets and super-Earths. The proximity of these of low-mass planets in the Solar Neighborhood will facilitate follow-up studies that would be difficult or impossible with \kepler\ stars, and their physical properties will inform the search for life outside of the Solar System.  Indeed, a small fraction of these nearby planetary systems have already been discovered \citep{Bonfils2007,55Cnce,Dumusque2012,Wittenmyer2014}.

Exoplanet transit searches require a near-perfect alignment of the orbital plane along the line of sight, necessitating fairly large sample sizes to ensure a detection. For example, the transit probability of a super-Earth orbiting a Sun-like star in an Earth-like orbit is approximately 0.5\%. This means a sample size of at least several hundred would be needed to ensure a single detection of one such planet. The limited number of local stars thus makes this method unfavorable for discovering large numbers of nearby planets. 

RV surveys are more promising, as the detection probability is less sensitive to orbital inclination. However, the velocity precision required for detecting small planets is just at or beyond the limits of most current instruments. Equally important to the success of such an RV survey is the tremendous cadence and phase coverage needed to densely sample a planet's full orbit and a range of stellar noise sources \citep{Dumusque2012,Endl2014}. Attaining this cadence for more than a small handful of stars is not realistic within the framework of shared telescope time allocation. Expanding our planetary census to dozens of stars in the Solar Neighborhood requires a dedicated observatory capable of highly precise RV measurements.

Some fraction of nearby planets discovered via the RV method will transit their host star, with an increase in transit probability for planets in shorter orbital periods. The RV data can therefore be used to guide searches of the transit windows of low-mass planets with precise photometry. This RV-first, transit-second method has proven to be a powerful observational technique producing the detection of the first transiting planet in 2000 \citep[HD 209458\,b;][]{Henry2000,Charbonneau2000}, and providing us with the brightest transiting systems known \citep{Sato2005, Winn2011}. These bright ($V<8$) systems are important because they are the most amenable to follow-up science using space and ground-based facilities.

In addition to the short--period ($P \lesssim 10$~days) transiting planets around nearby stars, there should also exist a large population of super-Earths 
that lie within their respective ``Habitable Zones'' \citep{Kasting1993,Kopparapu2013}. There have been a few examples from this population recently discovered that support this claim \citep{Anglada-Escude2012,Wittenmyer2014,Robertson2014}. 
Locating and characterizing these planets from the ground will populate the target lists of future space-based missions designed to produce direct images and spectra of Earth-like planets. In this way, the high-precision RV surveys of today will be an important stepping stone toward discovering a true analog to our own Earth.

We aim to address the need for new Doppler-based planet detection facilities by building a dedicated ground-based observatory for the detection of small planets in our Solar Neighborhood called the MINiature Exoplanet Radial Velocity Array (MINERVA). The philosophy and design of MINERVA are presented in Section~\ref{sec:design} which gives an overview of the project concept followed by a recent status update. The basic hardware that will be used to carry out MINERVA science is described in Section~\ref{sec:hardware} which includes the telescopes, their enclosures and the photometry cameras. A brief presentation of the MINERVA spectrograph is presented in this section; the design details and on-sky performance will be presented in a forthcoming publication. The software that will run the MINERVA array is being adapted from the robotic brain of the Robo-AO system \citep{Riddle2012} and we summarize its basic architecture and functionality in section \ref{sec:software}. In Section \ref{sec:commissioning} we present results validating the expected performance of our telescopes and cameras with photometric observations of one of our RV survey target stars, and we also present new observations and model fits to the transiting ``hot-Jupiter" WASP-52\,b in Section~\ref{sec:science}. Lastly, in Section~\ref{sec:conclusions} we briefly summarize this publication and offer a look toward future MINERVA opportunities.

\section{A Dedicated Facility for Exoplanet Detection and Characterization}
\label{sec:design}
MINERVA is designed around the primary science goal to detect super-Earths within the Habitable Zones of nearby stars, as well as terrestrial-mass planets in close-in orbits. For this latter class, the planet candidate orbiting $\alpha$~Cen\,B is the prototype \citep{Dumusque2012}. Our strategy to achieve this goal is to build a dedicated observatory and perform a precise radial velocity survey with a nominal 3 year time baseline. We will target solar-type stars in the optical part of the spectrum based on optimizations performed by our team\citep{Bottom2013} and the expertise of the project Principle Investigators. The relatively small number of targets, their brightness, their distribution on the sky, and our signal-to-noise ratio requirements to achieve our target precision of 0.8\,\ms\ along with estimated cost and time line for construction are the primary considerations that determine the nature of the telescope(s) required for this program. Multiplexing is not feasible since the stars will, in general, be widely spaced on the sky, and to achieve our target RV precision it would be inadvisable to risk crosstalk between multiple stellar spectra in a single exposure.

Large aperture telescopes ($D \gtrsim 3$\,m) have a relatively low duty cycle for these bright stars ($3 \lesssim V \lesssim 8$) because exposure times are short compared to the overheads incurred from CCD readout, slew time, and source acquisition. For example, the exposure time needed to achieve optimal signal to noise per pixel for a star with $V=5$ using the HIRES spectrometer on the Keck~I 10\,m telescope is about 5--10 seconds while the readout time is 42 seconds and a typical source acquisition time is 1--2 minutes. In addition, mitigating the effects of p-mode oscillations is best accomplished by matching exposure times to multiple oscillation periods, which in these stars is of order minutes \citep{Dumusque2011a}. Indeed, early radial-velocity planet searches worked effectively with apertures from 0.6--3 meters (two prominent examples are described in refs. \citenum{Marcy1996,ELODIE}).

\begin{figure}
\begin{center}
\begin{tabular}{c}
\includegraphics[width=\columnwidth]{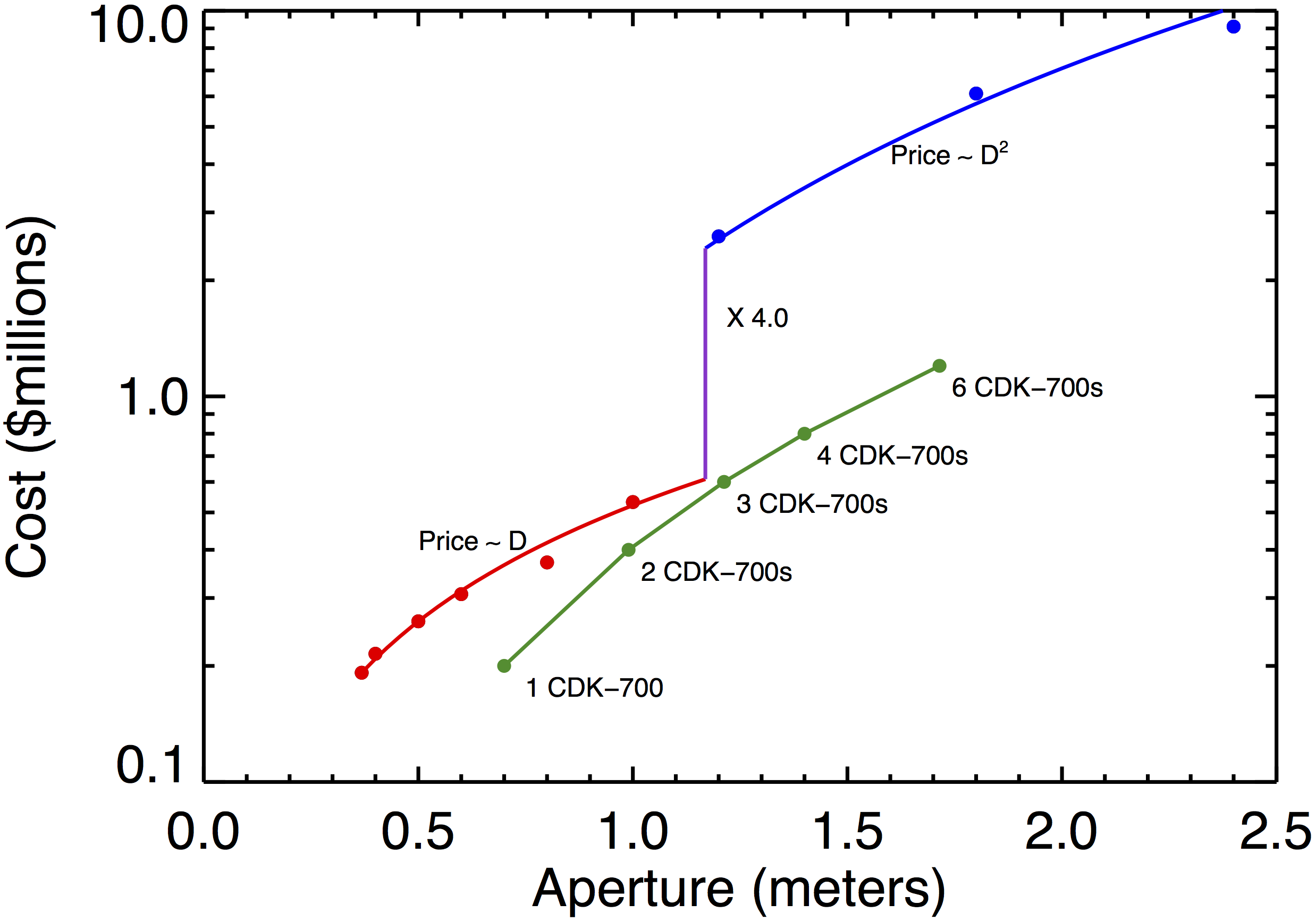}
\end{tabular}
\end{center}
\caption{\label{fig:cost}Cost curve for the purchase of telescopes of a specified aperture. The cost of amateur telescopes seems to scale with aperture diameter ({\it red points and curve}). There is a discontinuity near an aperture of 1\,m between the largest commercial telescopes and slightly larger professional telescopes that seem to follow a cost curve that scales as collecting area ({\it blue points and line}). The cost and effective apertures corresponding to successive numbers of CDK-700 0.7\,m telescopes are shown in green.} 
\end{figure}

Another benefit to using small telescopes is cost. We have conducted an informal survey of telescope costs by obtaining list prices for high-end, commercially-available small telescopes and cost estimates for larger, professional, custom telescopes up to 2.4 meters through private consultation\footnote{survey was conducted circa 2010 and may not reflect the current state of the market}. The data shown in Figure~\ref{fig:cost} reveals that the cost of small telescopes scales roughly as the aperture. There is a significant discontinuity in this scaling, however, between the largest commercial telescopes (roughly 1 meter) and slightly larger professional telescopes, of about a factor of four, presumably reflecting the economies of scale and less stringent engineering requirements of the amateur market.  Interestingly, extrapolation of the ``custom'' scaling to very large telescopes approximately describes the  cost of the Keck 10-m telescopes ($\sim$ \$100 million each) and the Thirty-Meter Telescope (TMT, $\sim$ \$1 billion). We also note that there appears to be a separate, lower track for truly mass-produced small telescopes (not shown in figure).

Since cost scales as aperture, there is no reduction in hardware cost in buying a single telescope vs.\ several small telescopes. We can take advantage of this fact to put MINERVA's cost onto the high-end amateur track by using multiple commercial 0.7\,m telescopes, whose light can be combined to create a larger effective aperture. This allows us to construct a large ``light bucket'' by feeding our spectrograph with multiple, smaller light buckets. The exact model of our 0.7\,m telescopes was chosen based on the specific features offered such as two instrument ports per telescope and fast slew speed as well as the proximity of the manufacturer to our test facility (see Section~\ref{sec:telescopes}).

In addition to the factor of four decrease in hardware cost over a single custom telescope, this design choice offers several advantages over using a single telescope. They can be purchased off-the-shelf, complete with control software, allowing a quicker path to on-sky commissioning and diminished development risks, which offset the increased complexity of organizing a telescope array.  The smaller \'{e}tendue of the optical systems translates to a smaller spectrograph that is easier to stabilize both mechanically and thermally. This reduces the cost of both the spectrograph and the facility needed to adequately stabilize the instrument.  Lastly, the modularity of the MINERVA design offers several benefits such as redundancy, the ability to change the scope of the project, and flexibility with our observing strategy.

\begin{figure}
\begin{center}
\begin{tabular}{c}
\includegraphics[width=0.99\columnwidth]{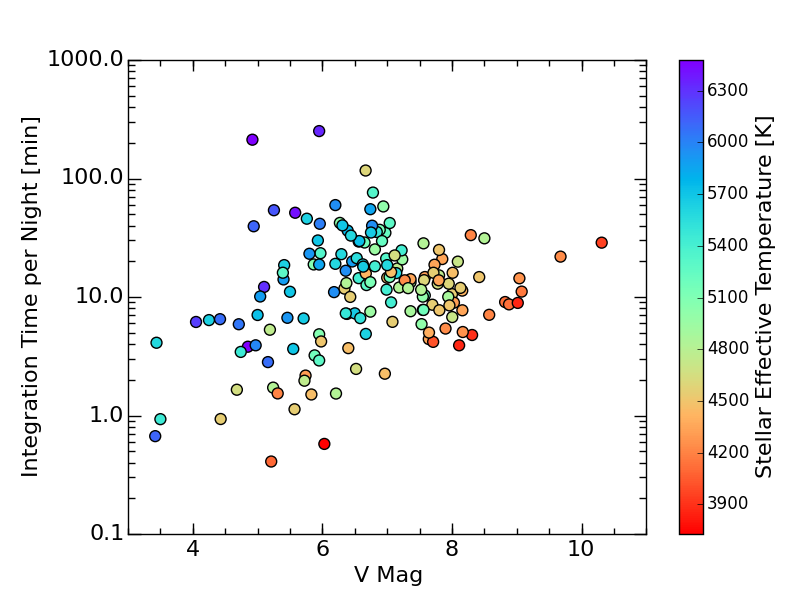}
\end{tabular}
\end{center}
\caption{\label{fig:sample}The required integration time per night for the MINERVA array to detect 3\,\mearth\ planets within the habitable zones of each star in the $\eta_\oplus$ sample according to a photon limited noise model as a function of $V$ magnitude. Data points are colored according to their effective temperatures, determined using stellar masses from ref.~\citenum{Howard2010b}.} 
\end{figure}

The survey target list is drawn from the {\em NASA/UC $\eta_\oplus$} sample made up of 166 nearby, chromospherically inactive stars currently monitored by Keck/HIRES for orbiting exoplanets \citep{Howard2009}. The projected yield from this target list dictates the minimum effective aperture required for MINERVA. We choose from this list the maximum number of targets that can be observed from the final location of the array in southern Arizona to the precision necessary to detect planets with \msini $ = $3\,\mearth\ in their respective habitable zones with 3 years of observations \citep{McCrady2014}. Figure~\ref{fig:sample} shows the $\eta_\oplus$ target list with the required integration time per night. As we do not have commissioning data for our spectrograph in hand, we perform simulations assuming a hard RV precision limit of 0.8\,\ms\ and a photon noise model. We assume a total system throughput of 10\% (see Sections~\ref{sec:spectrograph}, \ref{sec:telthroughput}, and \ref{sec:fiberthroughput}).

Our simulations account for stellar jitter, target observability, and weather losses based on historical weather records for southern Arizona. The radial velocity stellar jitter is modelled to match the spot-induced photometric variations observed in quiet G and K dwarfs by \kepler\ \citep{Basri2011}. We find that a spot model with 4 to 7 spot pairs per star with sizes ranging from 1.4\% to 1.8\% of the stellar radius reproduce well the observed photometric variations of between $3\times10^{-4}$ to $7\times10^{-4}$. We populate the surfaces of our simulated stars with spots having lifetimes and latitudes following the Solar “butterfly diagram” and assign rotation periods based on the period distribution seen by \kepler\ \citep{McQuillan2013}. The spot-induced pseudo-Doppler shifts are added to the simulated dynamical shifts caused by the planet. We then add white noise to the simulated radial velocities, recording the reduction in detectability of the simulated planet. The maximum amount of white noise added such that the planet was detected in 99\% of the realizations was taken as the required per-night precision per star. Our final target list is chosen in consideration of the length of nights, declination of target stars, calibration observations, 10\% overhead for secondary science and other programs, and a 25\,s telescope slew and source acquisition time.
 
The projected exoplanet yield from the MINERVA Doppler survey is estimated using the statistical results from the \kepler\ Mission \citep{Howard2012}. We extrapolated the reported occurrence rates out to periods of 400 days for planets above 2\,\rearth\ and assume the same frequencies per log bin as those for 85-day periods. The latter is likely a conservative assumption, as the frequency of planets appears to rise beyond 50-day periods. For each target in the MINERVA sample, we randomly drew planets in the radius-period grid based on the extrapolated frequency surface and then converted the exoplanet radius to a mass using a density relation, $M \propto R^{2.29}$ \citep{Howard2012}. Signals larger than 3\,\ms\ would have already been detected by the $\eta_\oplus$ program, and are thus not included in the yield. Multi-planet extractions were not performed in this simulation and may delay confirmation of the largest RV signals in multi-planet systems.

The final yield results are obtained from the results of repeating the simulation 1000 times. Using an effective aperture of 1.4\,m (4 CDK-700 telescopes) we are able to observe the 82 brightest stars from the $\eta_\oplus$ list, and we find a mean yield of $15\pm4$ planets with amplitudes between $0.8$\,\ms\ and $3$\,\ms. Improvements to these estimates are currently being pursued with the use of a more realistic stellar activity model and optimized observing strategies based on our recovery methods. Modeling of the stellar activity cycles will be necessary for the lowest mass planets and we recognize that the efficacy of these algorithms, which have yet to be quantified, will effect our estimated yield.

However, the ability to measure Doppler shifts of our targets at this precision every night (weather permitting) is unique to MINERVA and will be a key factor in recovering the RV signals of low-mass planets in their respective habitable zones. The observing cadence achievable with MINERVA allows us to account for the stellar variability of our sample of stars in a way that current facilities cannot. Based on our simulations and our expectations from a more refined treatment of stellar variability we take a 1.4\,m effective aperture to be the minimum required for ensured success of the project. We therefore design MINERVA around the use of four 0.7\,m telescopes.

The light from each telescope will be fed into 50\,\micron\ octagonal fibers using a custom focal plane unit [see \cite{Bottom2014} and Section~\ref{sec:fiber}]. These four fibers will then form a pseudo-slit at the entrance of the MINERVA spectrograph. The highly stabilized, bench mounted spectrograph will cover wavelengths from 500\,nm to 630\,nm with a resolving power of $R \approx 80,000$ (see Section~\ref{sec:spectrograph}). This is an optimal spectral range for RV surveys of solar-type stars \citep{Bottom2013} that can also be wavelength calibrated with an iodine cell \citep{Marcy1992}.

Built in to the design of MINERVA is the capability for flexible scheduling and simultaneous science and education programs. Of the $15 \pm 4$ simulated detections described above, $1.0\pm0.8$ are expected to transit their host star. Currently there are 16 RV detected planets with declinations $\delta > -20^\circ$, periods less than 30 days, $M\sin i < 50$\,\mearth\ and $V < 10$. With an additional $\sim 10$ from MINERVA, the total transit yield is expected to exceed unity. This motivates the secondary science objective of MINERVA to search for transits of super-Earths among its RV discovered planets and to further characterize known transiting planets with multi-band light curves. This requires a broadband photometry precision of $<1$\,mmag in the optical. We demonstrate a comparable level of photometric precision from our commissioning site on the Caltech campus in Pasadena, CA (see Section~\ref{sec:commissioning}). On the educational front, students in lab courses can use one of the telescopes to conduct their course assignment and gain valuable observing experience---a community need recently expressed by \cite{Tuttle14}---while the other telescopes simultaneously conduct the primary science program.
		
During the early stages of the project MINERVA will be used to follow up newly found Jupiter- and Neptune-sized planets from surveys like HATNet \citep{Bakos2002}, WASP \citep{Pollacco2006} and NGTS \citep{Wheatley2013} in addition to providing long-term monitoring for some TERMS long period planets \citep{Dragomir2011}. MINERVA photometry can also be used to follow up space-based discoveries. The \kepler\ K2 mission \citep{Howell2014} will produce thousands of transit discoveries, but will only monitor each target field for approximately 75 days. At later times the number of potential targets will grow considerably as the TESS Mission \citep{Ricker2014}, set to launch in 2017, will yield many more detections over the whole sky, but with continuous monitoring of only 27 days for each non-overlapping field.

\subsection{Project Status and Approximate Timeline}
\label{sec:status}
As of the writing of this manuscript, MINERVA telescopes 1 and 2 are located on the Caltech campus in the first ``Aqawan'' (see Section~\ref{sec:aqawan}) where we have been developing the MINERVA Robotic Software (MRS) and validating the performance of the various hardware components. The hardware on-site has been fully tested, and now the goal of this facility is to achieve coordinated, automated control of both telescopes and Aqawan 1. This is expected to be complete by early 2015, and once this goal has been reached the entire facility will be moved to Mt. Hopkins. 

The infrastructure at Mt. Hopkins necessary for the relocation of the Aqawans and telescopes has been completed. Aqawan 2 has been constructed in California and was transported to Mt. Hopkins on December 9, 2014. MINERVA telescopes 3 and 4 have had their performance validated and were relocated to Aqawan 2 at Mt. Hopkins on December 15, 2014.  

The custom room designed for the spectrograph is currently under construction. The outermost layer in a three stage environmental control scheme will be a 100K clean room temperature stabilized to $\pm 1^\circ$C, peak-to-peak. A second, interior room will then be erected inside which the spectrograph will be mounted with its critical optical elements inside a vacuum chamber. The manufacturing of the spectrograph has been completed by Callaghan Innovation in New Zealand. Work is now underway on the input optics as well as the iodine cell mount. It is undergoing lab tests and awaits the completion of the spectrograph room at Mt. Hopkins. Delivery of the spectrograph is expected by the end of the first quarter of 2015 when on-sky commissioning will begin. 

Allowing for minor unforeseen delays, fully automated, robotic control of the array and spectrograph is expected by mid-year 2015. The primary survey is projected to begin by the end of the year.

\section{MINERVA Hardware}
\label{sec:hardware}
\subsection{The CDK-700 by PlaneWave}
\label{sec:telescopes}

The PlaneWave CDK-700 is a 0.7 meter, altitude/azimuth mounted telescope system\footnote{\url{http://planewave.com/products-page/cdk700}} \citep{Hedrick2010}. It has a compact design, standing just under 8 feet tall when pointing at the zenith, with a 5 foot radius of maximum extent when pointing horizontally. The telescopes use a corrected Dall-Kirkham optical setup consisting of an elliptical primary mirror, a spherical secondary mirror, and a pair of correcting lenses to remove off-axis coma, astigmatism, and field curvature. This results in a flatter, more coma and astigmatism-free field than the Ritchey-Chretien design, with the added benefit that the spherical secondary mirror makes alignment forgiving compared to the hyperbolic secondary of the Ritchey-Chretien design. The CDK-700 has dual Nasmyth ports with output beams at f/6.5 accessed with a rotating tertiary mirror. The CDK-700 specifications are summarized in Table~\ref{tab:telescope}.

\begin{center}
\begin{deluxetable}{p{3cm}p{5cm}}
\tablewidth{\columnwidth}
\tablecaption{CDK-700 Specifications}
\startdata
\hline \\
\hline \\
\multicolumn{2}{c}{\bf{Optical System}} \\
\hline \\
Optical Design\dotfill        & Corrected Dall-Kirkham (CDK) \\ 
Aperture\dotfill              & 700\,mm (27.56 \,in) \\
Focal Length\dotfill          & 4540\,mm \\
Focal Ratio\dotfill           & 6.5  \\
Central Obscuration\dotfill   & 47\% primary diameter \\
Back Focus\dotfill            & 305\,mm from mounting surface \\
Focus Position\dotfill        & Nasmyth (dual) \\
Dimensions\dotfill            & $93.73^{\prime\prime}$\,H $\times$ $43.25^{\prime\prime}$\,W $\times$ $39^{\prime\prime}$\,D \\
Weight\dotfill                & 1200\,lbs \\
Optical Performance\dotfill   & 1.8\,$\mu$m RMS spot size on axis \\
Image Scale\dotfill           & 22\,$\mu$m per arcsecond \\
Optimal Field of View\dotfill & 70\,mm (0.86 degrees) \\
Fully Baffled Field\dotfill   & 60\,mm \\
\cutinhead{\bf{Mechanical Structure}}
Mount\dotfill                 & Altitude-Azimuth \\
Fork\dotfill                  & Monolithic U-shaped fork arm \\
Azimuth Bearing\dotfill       & 20\,in diameter thrust bearing \\
Altitude Bearing\dotfill       & $2 \times 8.5$\,in OD ball bearings \\
Optical Tube\dotfill           & Dual truss structure \\
\cutinhead{\bf{Motion Control}}
Motors\dotfill                 & Direct drive, 3 phase axial flux torque motor \\
Encoders\dotfill               & Stainless steel encoder tape with 81\,mas resolution \\
Motor Torque\dotfill           & $\sim 35$\,ft-lbs \\
Slew Rate\dotfill              & $15^\circ$\,s$^{-1}$ \\
\cutinhead{\bf{System Performance}}
Pointing Accuracy\dotfill      & $10^{\prime\prime}$ RMS \\
Pointing Precision\dotfill     & $2^{\prime\prime}$ RMS \\
Tracking Accuracy\dotfill      & $1^{\prime\prime}$ RMS over 3 minutes \\
Field De-Rotator\dotfill       & $3\,\mu$m peak-to-peak 35\,mm off axis over 1\,hr \\
\enddata 
\label{tab:telescope}
\end{deluxetable}
\end{center}

The telescope pointing is controlled by two direct--drive electromagnetic motors with encoder resolution of 81\,mas resulting in a pointing accuracy of $10^{\prime\prime}$ RMS, a pointing precision of $2^{\prime\prime}$, and a tracking accuracy of $1^{\prime\prime}$ over a three-minute period. The slew rate is $15^\circ$\,sec$^{-1}$ which keeps slew times between any two points on the sky to less than $\sim 10$ seconds. The focus mechanism and image de-rotator are combined into a single, motor-controlled unit that can be remotely adjusted. Cooling fans and temperature sensors are used to keep the primary mirror in thermal equilibrium, and the control software is built to automatically correct for perturbations such as wind gusts.

\begin{figure}
\includegraphics[width=\columnwidth]{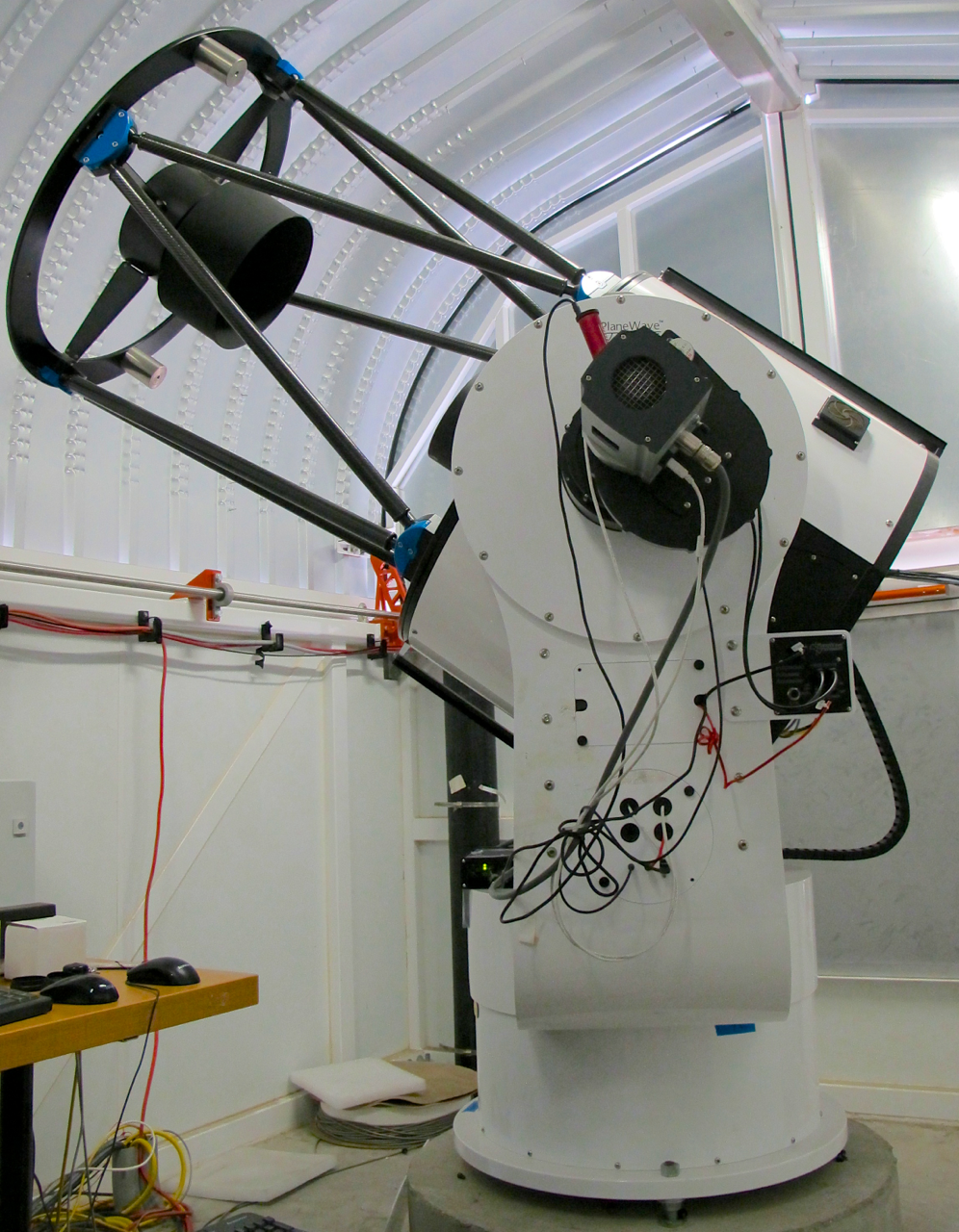}
\caption{\label{fig:cdk700}Telescope 2, a PlaneWave CDK-700, is shown inside the MINERVA Aqawan telescope enclosure at the Caltech commissioning site.} 
\end{figure}

Section~\ref{sec:commissioning} presents the performance validation for our telescopes including control, pointing, guiding, and throughput. The commissioning performed over the past 2 years at Caltech has proven the feasibility of using high-end, off-the-shelf hardware for professional astronomical research. Thereby, MINERVA has retired many of the risks involved in developing an observational facility from the ground up. Figure~\ref{fig:cdk700} shows an image of the MINERVA telescopes at the Caltech test site.

\subsection{The ``Aqawan'' Telescope Enclosures}
\label{sec:aqawan}
\begin{figure}
\includegraphics[width=\columnwidth]{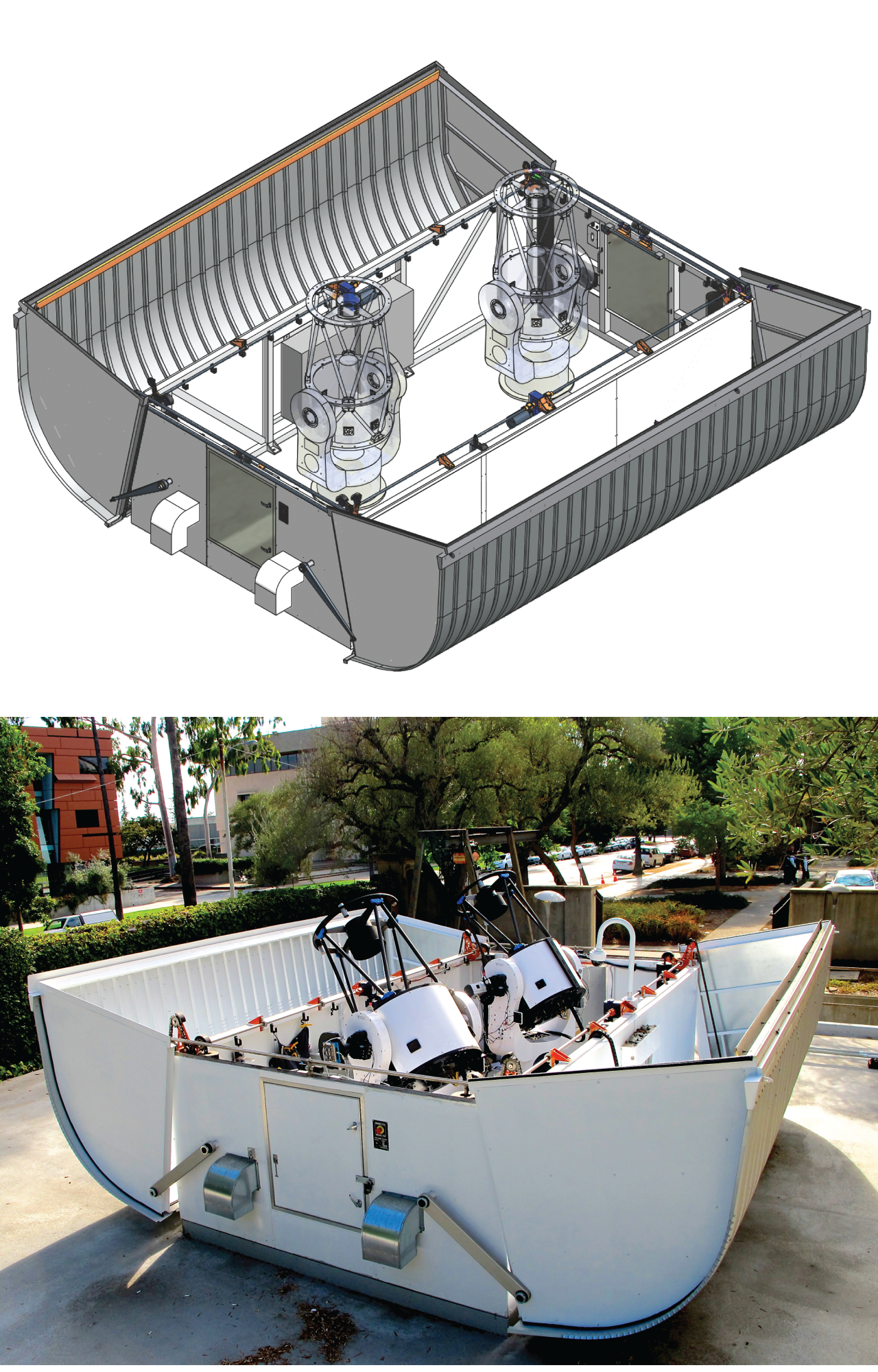}
\caption{\label{fig:aqawan}({\it left}) Design drawing of our custom Aqawan telescope enclosure with two PlaneWave CDK700 telescopes inside. ({\it right}) MINERVA commissioning site on the Caltech campus showing the open Aqawan and telescopes 1 and 2 inside.} 
\end{figure}

An ``Aqawan'' (Chumash native American word for ``to be dry'')  is a telescope enclosure developed by Las Cumbres Observatory Global Telescope (LCOGT) for their 0.4\,m telescopes specifically designed for remote, robotic operations around the world \citep{Brown2013}. The design offers full access to the sky, limiting the effects of ``dome seeing'', and eliminates the need to coordinate dome slit positioning while maintaining a relatively small footprint. We have purchased two custom--built Aqawans with longer sides that can each accommodate two CDK-700s without any possibility of collision, and can close safely with the telescopes pointed in any orientation. Stronger motors with a higher gear ratio were also installed to handle the heavier roof panels. Figure~\ref{fig:aqawan} shows the design and realization of our first Aqawan which has been delivered to the Caltech campus for commissioning.

The Aqawan receives 208V/30A three phase power that is converted to 24V DC within its control panel. This power runs through an internal uninterruptible power supply (UPS) and powers a programmable automated controller (PAC) that controls all the functionality of the enclosure. In addition to basic opening and closing, the Aqawan has many auxiliary features including: a web camera that provides 360$^\circ$ coverage, a temperature and humidity sensor, fans to promote temperature equalization, fluorescent lighting, and a smoke alarm. Communication to the Aqawan PAC is established via TCP/IP, with commands consisting of ASCII strings. The Aqawan firmware is designed such that if a ``heartbeat" command is not issued each minute, the roof automatically closes. This feature along with built-in backup power and the ability for the Aqawan to close with the telescopes in any configuration offers safety against power and connectivity failure modes.

\subsection{Fiber and Fiber Coupling}
\label{sec:fiber}
Astronomical light collected from MINERVA's four 0.7\,m telescopes will feed a stabilized spectrograph via fiber optic cables. Using a fiber feed offers many advantages and some challenges. Single mode fibers offer superior control of the instrumental profile. However, they are diffraction-limited by nature and require a high-performance adaptive optics system for efficient coupling of starlight. Multi-modal fibers couple to starlight much more easily, but they have a near and far-field output that is variable. The near-field variations are due to the interference of modes at the output of the fiber, and when imaged on the detector these variations can limit the signal to noise ratio of the observations. This effect can be mitigated by use of an octagonal fiber which mixes the modes as they traverse the fiber \citep{Chazelas2010,Bouchy2013}. Physical agitation of the fiber enhances this effect \citep{Baudrand2001}.

The far-field uniformity and stability are also important, as the far field is incident on the echelle grating.  Different grooves will be illuminated if the spatial intensity distribution changes, which can introduce spurious wavelength shifts as the grooves are not identical. One way to improve this performance is to introduce a double scrambler, which inverts the field and angle distribution, at a cost of reduced throughput \citep{Barnes2010}. This is an option that MINERVA will explore further if necessary.

We will use 50\,\micron\ octagonal fibers with a 94\,\micron\ cladding diameter and a numerical aperture of 0.22 to feed light collected from each of the four 0.7\,m telescopes to our stabilized spectrograph. Starlight will be coupled to our fibers at the native f/6.5 of the CDK-700 bypassing the need for small optics for the purpose of better throughput. The spectrograph is designed to accommodate an f/5.0 beam or larger, permitting up to 20\% focal ratio degradation before impacting overall throughput. Therefore our fibers project onto the sky with a $2.27^{\prime\prime}$ diameter. 

Our fiber coupling system consists of a fiber acquisition unit (FAU) and control software that can interface with the unit and provide closed loop guiding with the telescope control. Besides the MINERVA spectrograph, this is one of the only truly custom hardware components of MINERVA. The details of this system are presented by \cite{Bottom2014}, and here we present only a brief review.

\begin{figure}
\includegraphics[width=\columnwidth]{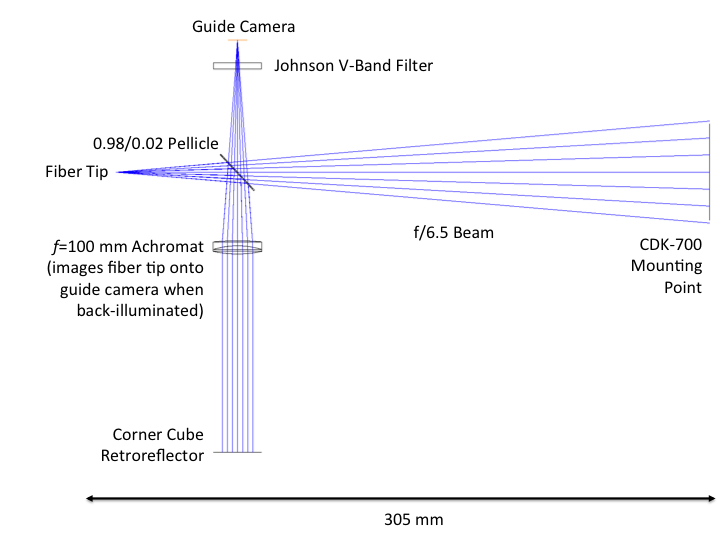}
\caption{\label{fig:FAUzemax}Ray-trace illustration of the fiber acquisition unit (FAU), which measures 305\,mm ($\sim$12 inches) across.  Light enters from the right, delivered by the CDK-700 with an f/6.5 beam.  A pellicle reflects 2\% of the beam to an SBIG ST-i guide camera with a V-band filter, matching the wavelengths used for radial velocity observations.  The remaining 98\% of the beam comes to a focus on the fiber tip.  We include an achromatic lens and corner cube on the opposite side of the pellicle such that the fiber tip can be imaged onto the guide camera via back illumination.  This allows for quickly determining the ``location'' of the fiber tip on the guide camera for precise guiding.} 
\end{figure}

\begin{figure}
\includegraphics[width=\columnwidth]{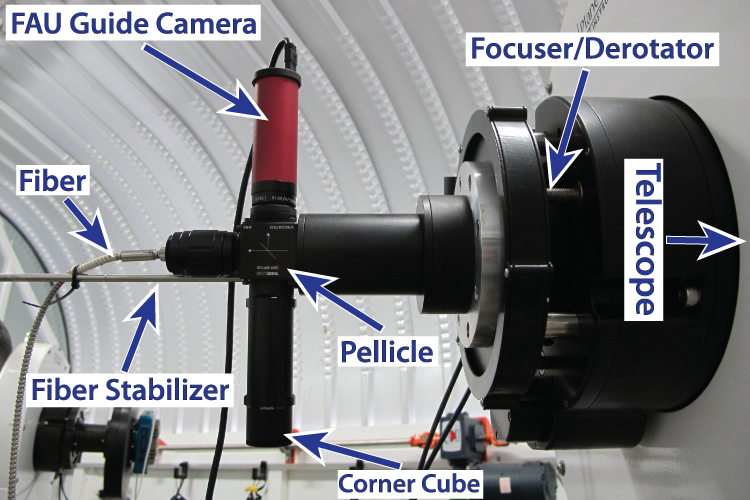}
\caption{\label{fig:FAUreal}The MINERVA FAU mounted on one of the MINERVA telescopes with all major components labeled. The focuser/derotator will be disengaged for standard spectroscopic observations.} 
\end{figure}

The FAU has three accessible optical paths that are based upon the design presented in by \cite{Plavchan2013}. In the primary path, the telescope beam is fed directly into the fiber. Separately, there is an optical path which relays a portion of the light to a guiding camera via a pellicle. Finally, there is a set of relay lenses and a corner retroreflector. If the fiber is illuminated from the exit end near the spectrograph, the input end of the fiber tip will be imaged on the guide camera. The corner retroreflector guarantees that misalignments of the optics do not affect the image position; this allows the determination of the pixel position on the guider that corresponds to the fiber tip, and hence a setpoint for guiding. An annotated schematic of the FAU design is shown in Figure~\ref{fig:FAUzemax}, and a picture of one of the FAUs mounted on a CDK-700 can be seen in Figure~\ref{fig:FAUreal}.


\subsection{The MINERVA Spectrograph}
\label{sec:spectrograph}
The MINERVA spectrograph is an adaptation of an existing spectrograph design \citep{Barnes2012a,Gibson2012} that leverages the successes of existing facilities \citep{Mayor2003,Cosentino2012,Crane2010} and techniques \citep{Marcy1992,Butler1996}. This critical component of our facility will be described in detail in a forthcoming publication. Here we provide a brief description of the spectrograph for the sake of completeness of this publication.

\begin{figure}
\includegraphics[width=\columnwidth]{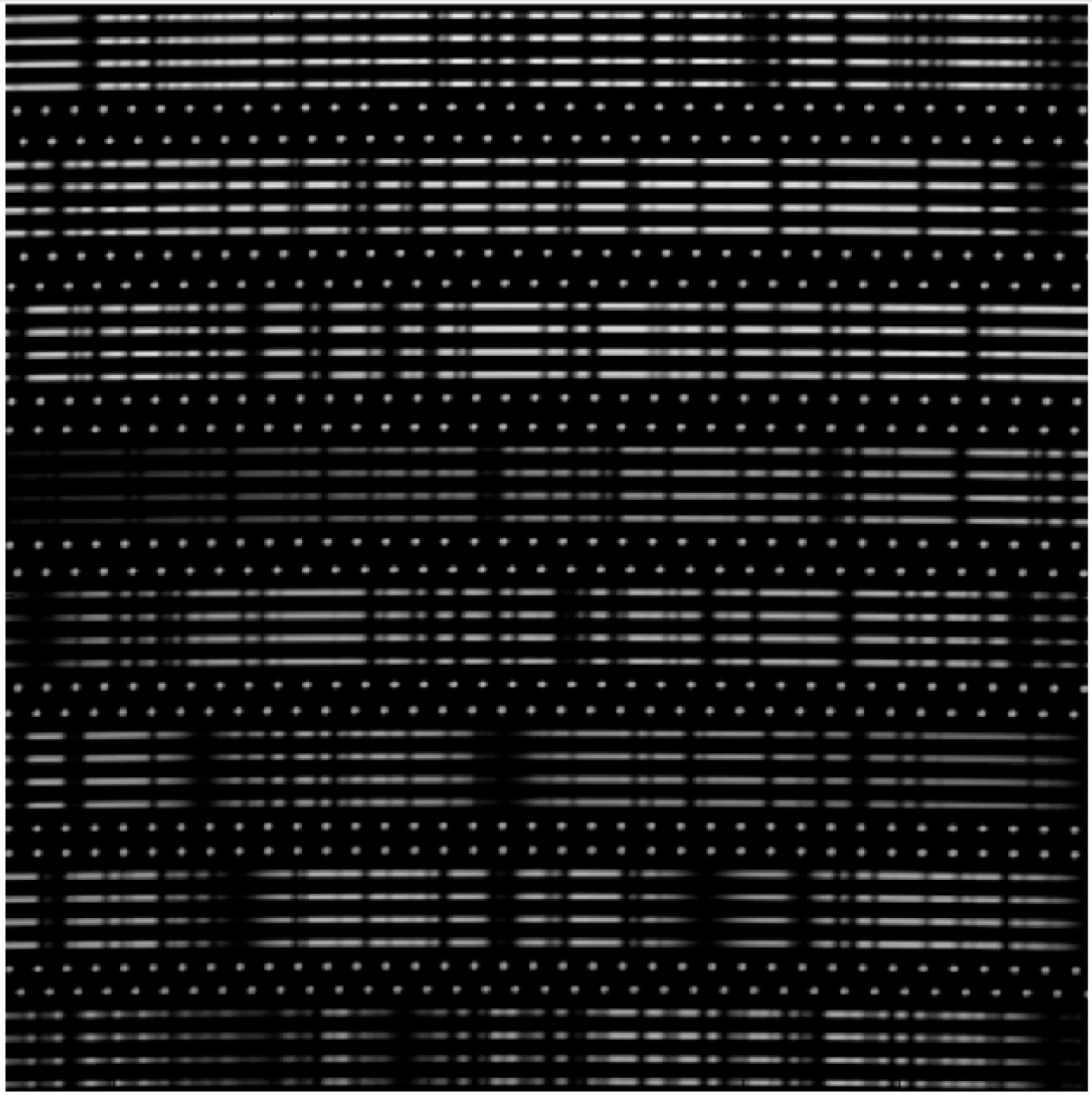}
\caption{\label{fig:echelle}A subsection of a simulated MINERVA echellogram showing four science spectra bracketed by a simulated stabilized etalon wavelength source. Each science fiber trace represents the input from one of the four MINERVA telescopes that will be observing the same astronomical target.} 
\end{figure}

It is a bench mounted, fiber-fed spectrograph of asymmetric, white pupil design. Primary dispersion is achieved using an R4 echelle, and a VPH grism is used for cross-dispersion. Four distinct traces will be imaged on a 2\,k$\times$2\,k detector covering a spectral range from 500 to 630\,nm over 26 echelle orders. The resolving power of the spectrograph is $R \approx 80,000$. Two additional calibration fibers bracket the 4 science fibers and provide stable wavelength calibration with the use of a stabilized etalon wavelength source (as simulated in Figure~\ref{fig:echelle}), or Thorium-Argon light. A subsection of a simulated echellogram created using ray-tracing techniques \citep{Barnes2012b} is presented in Figure~\ref{fig:echelle}. Each science trace corresponds to the input from an independent telescope and fiber system. 

The estimated throughput of the prototype spectrograph from which the MINERVA spectrograph has been designed is approximately $30$\%. This has been validated with on-sky measurements taken at Mt. John Observatory in New Zealand \citep{Gibson2012}. For these tests, the mirrors for the prototype spectrograph were bare aluminium, the lenses had only single layer anti-reflection coatings, and several off-the-shelf optical components were used, all of which contribute to a sub-optimal throughput. Improvements have now been implemented, including high-efficiency coatings, and the installation of a new custom camera such that the throughput is expected to increase to $\sim 45$\%. The throughput of our fiber system is expected to be $\approx 70$\%, and the total throughput of our system up to the entrance slit of the spectrograph is expected to be $\approx 50$\% (see Sections~\ref{sec:telthroughput} and \ref{sec:fiberthroughput}). Therefore the total throughput of our system including losses due to sky extinction at the Fred Lawrence Whipple Observatory is expected to be $10$\% or greater.

The line spread function will be sampled with 3 pixels and a minimum of 4 null pixels will lie between each trace. The spectral extraction and data reduction pipelines are currently being developed. The strategy is to model the 2D spectrum directly, {\ie}, all orders will be modeled simultaneously such that cross-contamination between orders and scattered light will be accounted for in the model. This approach is likely only possible in the high signal to noise regime in which MINERVA will be working. The fact that all telescopes will be observing the same target simultaneously will also help to mitigate the effects of cross-contamination between orders. 

The spectrograph will be placed inside a purpose built, 2-stage room and the critical components will reside inside a vacuum chamber stabilized to $\pm 0.01^\circ$C. An iodine cell will be mounted off the optical bench and can be inserted or removed from the optical path, allowing the option of simultaneous wavelength calibration of the science traces in the echellogram. 

\subsection{Cameras and filters}
\label{sec:cameras}
\begin{figure}
\includegraphics[width=\columnwidth]{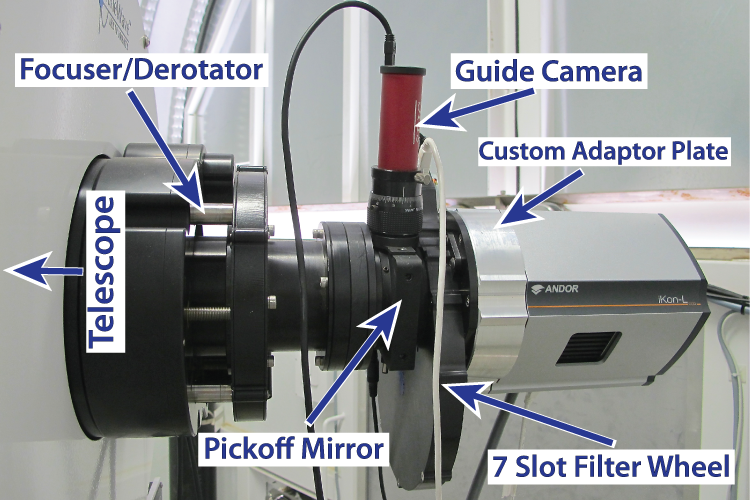}
\caption{\label{fig:camera}The optics chain for MINERVA photometry with major components along the light path to the camera labeled.} 
\end{figure}

Each MINERVA telescope will also be equipped with a wide field CCD camera on one of the two Nasmyth ports available on the CDK-700. The array will incorporate 3 different camera models for its 4 telescopes. Two of the MINERVA telescopes will be equipped with identical Andor iKON-L cameras.\footnote{\url{http://www.andor.com/scientific-cameras/ikon-ccd-camera-series/ikon-l-936}} These cameras have a back illuminated sensor with wide band (BV) coating and $2048\times2048$ square 13.5\,\micron\ pixels for a total chip size of 27.6\,mm corresponding to a $20.9^\prime$ field. Figure~\ref{fig:camera} shows an image of one of these cameras mounted on a MINERVA telescope. A third telescope will be equipped with an additional Andor iKON-L that is identical to the two described above except that it will contain a deep depletion sensor with fringe suppression (BR-DD). This camera is sensitive to light at wavelengths out to 1\,\micron\ and can be used for precision photometry in the near infrared ({\eg}, $i^\prime, z^\prime$, and $Y$). The iKON-L cameras come with a 4 stage thermo-electric cooling system that can achieve operating temperatures $\approx -80^\circ$C, nearly eliminating dark current. The third camera model will be an Apogee Aspen CG230\footnote{\url{http://www.ccd.com/aspen\_cg230.html}} with $2048\times2048 \ 15\times15$\,\micron\ pixels for a chip size of 30.7\,mm corresponding to a $23.2^\prime$ field.

Each MINERVA telescope has a filter wheel on the photometry port, and we employ two different Apogee filter wheel models, each with 50\,mm square slots. Three of the MINERVA telescopes have the AFW50-7s filter wheel while one telescope has a custom double filter wheel comprising two AFW50-10s wheels mounted back to back. One open slot in each 10 slot wheel allows access to the other 18 slots. Currently, the filters available to the MINERVA system are the Johnson $U$, $B$, $V$, $R$, and $I$; second generation Sloan $g^\prime$, $r^\prime$, $i^\prime$, and $z^\prime$; narrow band H\,$\alpha$, [S{\sc ii}], [O{\sc iii}], and amateur filters, $L$, $R$, $G$, and $B$. The Andor cameras are mated to the filter wheels through a custom adaptor plate designed by P. Gardner of Caltech Optical Observatories and implemented by Andor.  

\section{The MINERVA Site: the Fred Lawrence Whipple Observatory (FLWO)}
\label{sec:site}
The performance validation and preliminary commissioning of all MINERVA components aside from the spectrograph have been done at the commissioning site on the Caltech campus in Pasadena, CA (see Figure~\ref{fig:aqawan}), or at the PlaneWave warehouse in Rancho Dominguez, CA. Once the primary commissioning tasks are complete and fully automated observations have been achieved, the entire facility will be moved to its final location at the Fred Lawrence Whipple Observatory (FLWO) on Mt. Hopkins outside of Amado, AZ (see Section~\ref{sec:status}). 

The FLWO site was chosen after a site selection study conducted over the summer of 2013. We accumulated historical data and visited with staff and personnel from Mt. Wilson Observatory in Los Angeles, CA, McDonald Observatory in Jeff Davis County, TX, and San Pedro M\'artir Observatory in Baja California, Mexico in addition to FLWO. Mt. Hopkins was determined to be the optimal choice for MINERVA based on good overall weather and seeing conditions, the existing infrastructure available for use, full-time support staff, and financial considerations in setting up and maintaining the site.

\begin{figure}
\includegraphics[width=\columnwidth]{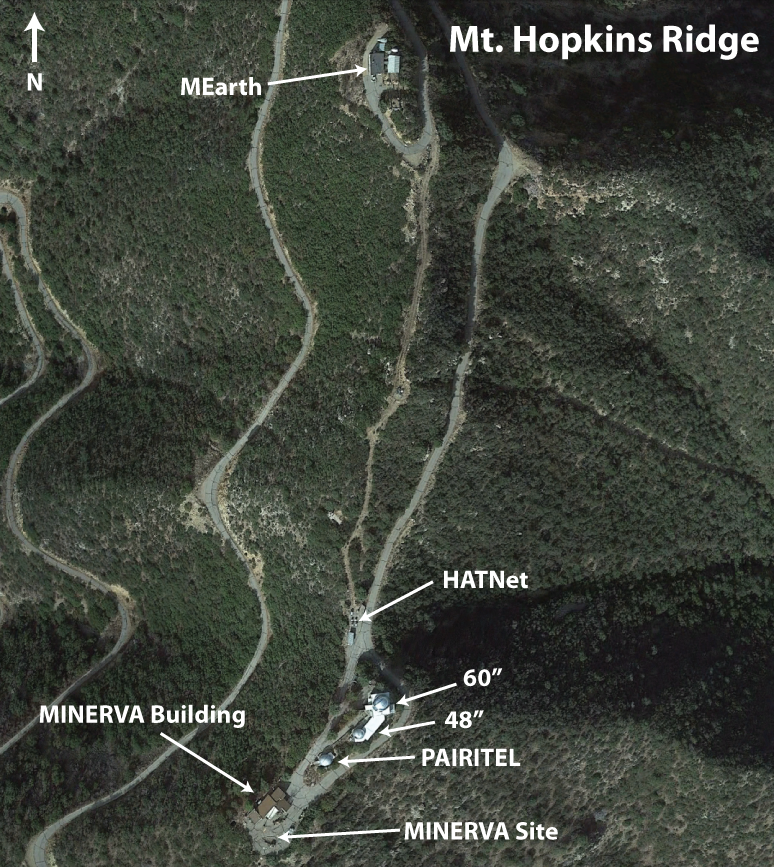}
\caption{\label{fig:ridge}Locations of current and future astronomical facilities on the Mt. Hopkins Ridge. The MINERVA site and building are labeled at the bottom of the image.} 
\end{figure}

The weather data for Mt. Hopkins consisted of a compilation of observing logs from the HAT-Net project \citep{Bakos2002} located on the Mt. Hopkins Ridge approximately 130\,m to the North, and data from the MEarth project \citep{Irwin2009} located about 430\,m to the North and at a slightly higher elevation than MINERVA (see Figure~\ref{fig:ridge}). We also incorporated information presented by \cite{JGibson2012} as the relative conditions at the Ridge correlate well with the conditions at the summit where the MMT is located. From these data we anticipate approximately 271 nights per year that we will be able to observe for 6.5 hours or more with a median seeing of $1.2^{\prime\prime}$. 

The MINERVA site is located at ($\phi$, $\lambda$) = $31^\circ$ $40^\prime$ $49.4^{\prime\prime}$ N, $110^\circ$ $52^\prime$ $44.6^{\prime\prime}$ W at an elevation of 7816 feet. Figure~\ref{fig:ridge} shows the MINERVA telescope and building location in relation to the rest of the astronomical facilities on the Mt. Hopkins Ridge. The MINERVA telescopes will all be placed on the flat and smooth area approximately where the decommissioned FLWO 10\,m gamma-ray telescope was located. The horizon limits have been measured to be between $15^\circ$ and $20^\circ$ over the full $360^\circ$ azimuth range except for an 80-foot lightning rod northwest of the array which covers a small total solid angle, but must be considered when tracking high declination sources across the sky, and the top corner of the MINERVA building which reaches an elevation of approximately $30^\circ$ north of the MINERVA telescope locations. The foundations for two Aqawans oriented with their long axes East-West are 2 feet thick reinforced concrete that have been properly grounded. The telescope piers are 36 inches in diameter and extend approximately 5 feet below the ground surface. One of the piers is anchored into the underground remains of the 10\,m pier. 

MINERVA will use the building that was used for the FLWO 10\,m gamma ray telescope that is now decommissioned and whose function was replaced by VERITAS\footnote{\url{http://veritas.sao.arizona.edu/}}, located at the base of Mt. Hopkins. There are several rooms in the building, three of which are instrumental to the operation of MINERVA: 1) the spectrograph room is the southeastern-most room that is being converted into a class 100,000 clean room within which the MINERVA spectrograph will be mounted; 2) the UPS room is located immediately northwest of the spectrograph room where the power for the entire MINERVA facility will be routed through a facility grade uninterruptible power supply (UPS); and 3) the control room adjacent to the spectrograph room, which will house control computers and network equipment.

\section{MINERVA Robotic Software (MRS)}
\label{sec:software}
\begin{figure*}
\begin{center}
\includegraphics[width=6.5in]{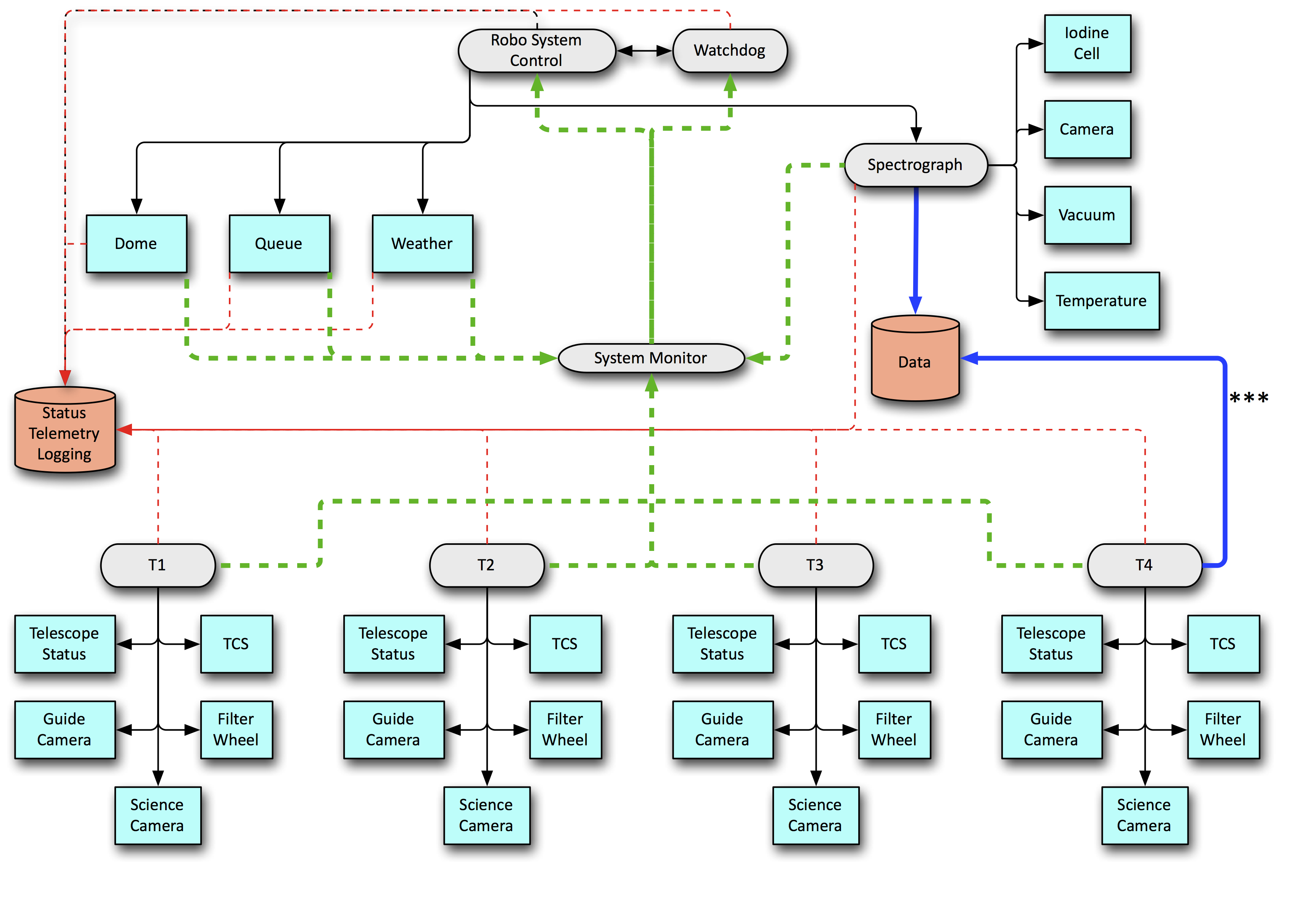}
\caption{\label{fig:software}MINERVA Robotic Software (MRS) architecture. Blue boxes are subsystem control daemons, gray boxes are control or oversight daemons that control more than one daemon, and red boxes represent data file storage. The red dashed lines with arrows signify the paths for telemetry through the system, black lines denote the command paths, and the blue are the data paths (the three asterisks indicate three other data pathways from T1, T2, and T3 that we leave off for the sake of neatness). The green lines signify the pathways for the status information that is passed to the system monitoring software.} 
\end{center}
\end{figure*}

MINERVA will be a completely autonomous facility. The MINERVA Robotic Software (MRS) is being adapted from the Robo-AO software \citep{Riddle2012} that has been successfully operating with a laser guide star adaptive optics system on the 60-inch telescope at Palomar Observatory for the past 3 years. Robo-AO has already completed the largest adaptive optics surveys to date with high observing efficiency and robust operation \citep{Baranec2014}. The Robo-AO software was developed in a modular way such that it can be easily replicated and used for the robotic operation of laser adaptive optics imaging on other 1 to 2\,m class telescopes. However, this design also allows a straightforward adaptation of the Robo-AO brain to the MINERVA system. 
The MRS architecture is shown in Figure~\ref{fig:software}. The operation of MINERVA is carried out by 6 separate computers, {\sc main}, {\sc telcom1}, {\sc telcom2}, {\sc telcom3}, {\sc telcom4}, and {\sc spec}. All computers use Ubuntu 12.04.2 as the base operating system, and all source code is written in C++. Communications between the subsystems of MRS use the custom TCP/IP protocol developed for Robo-AO. This protocol is used to pass commands and exchange telemetry between each of the subsystems. This system will detect when one of the control daemons for a subsystem dies and will restart the subsystem automatically. 

The main computer, {\sc main}, coordinates operation of the facility while monitoring the performance and state of all system components as well as environmental conditions. The three control or oversight daemons running on this computer are the Robo System Control, the Watchdog, and the System Monitor. The Robo System Control provides the high level control of MINERVA, coordinating the execution of all operations.

The System Monitor manages the information flow of the status of the subsystems to the entire robotic system. This part of the software regularly examines the status of each of the other software elements for their state of operation. It detects when one of the software subsystems has an error, crashes, or has other problems that might hinder the proper operation of the system. Issues are flagged and stop the operation of the automated system observations until the subsystem daemon can clear the issue. If the subsystem cannot correct the error, the automation system can take steps, up to and including restarting subsystems, in an attempt to continue operations. If it is unable to restart the system a message will be sent for human assistance, and an attempt can be made to continue operation without the failed subsystem, {\eg}, if one telescope fails the other 3 telescopes can continue with the primary science program. If operations cannot continue without the failed subsystem ({\eg}, the spectrograph), the software will close the Aqawans and shut everything down, leaving the system in a safe state.

The sole function of the Watchdog system will be to make sure the observing system remains in operation. If any system crashes or stops working properly, it will attempt to restart it, as well as stop telescope operations. If the process cannot be restarted and is not essential the system will continue without the failed subsystem. If the failed subsystem is essential, the shutdown sequence will be initiated followed by a request for human help.

Each of the four telescopes and their suite of instruments are run with a single computer on which the station daemons are run, T1 through T4. Each station daemon reports telemetry information and handles the operation of the subsystems mounted on each telescope including the telescope control system (TCS), guide cameras, imaging science camera, and filter wheel.

The {\sc spec} computer is dedicated to the control of the spectrograph, and runs the Spectrograph daemon which coordinates the operation of the spectrograph and monitoring of the spectrograph environment. This daemon also gathers data from the System Monitor and outputs the science data with extensive header information regarding the state of the facility during the time over which data were collected.

The software control of each hardware subsystem consists of a set of individual modules. Each interface module handles configuration file interactions, initialization, and error control. These modules are stacked together into larger modules, which are then managed by other facets of the robotic control system. The subsystems are run as daemons in the operating system; each separately manages the hardware under its control and runs a status monitor to sample subsystem performance and register errors that occur. Each of these subsystems is composed of many separate functions that initialize the hardware, monitor its function and manage the operation of the hardware to achieve successful scientific output. In essence, each of the subsystem daemons are individual robotic programs that manage their hardware and operate according to external commands. The subsystem daemons communicate their state through the TCP/IP protocol to the System Monitor, and the System Monitor then parses and relays this information to the Watchdog and Robo System Control.

MRS will also employ a queue system modeled after the Robo-AO system built to read and organize program and target files in XML format \citep{Riddle2014}. While the elimination and weighting criteria for the MINERVA program will be slightly different than for Robo-AO, the approach will be essentially the same for the primary MINERVA spectroscopic survey. MINERVA will operate primarily in a coordinated fashion. Targets will be observed by all telescopes simultaneously, vastly simplifying operations. On the occasion when one or more of the telescopes is needed for another program, {\eg} transit photometry, a natural break in operations will be identified when the needed telescope(s) can be seamlessly left out of the next observing sequence. The telescope(s) will then be released from robotic control and can be operated remotely.

Development of a robotic system from scratch is an involved process, and can take years, especially for a system controlling multiple enclosures and telescopes.  Using the Robo-AO software as a base, the MINERVA software development time has been cut by at least 50\%.  Most subsystems are under computer control and debugged; work continues on the TCS interface, and detailed spectrograph software development is awaiting delivery of the spectrograph hardware.  Software to control the individual stations and the overall robotic software control system (including the queue scheduler) is currently undergoing final development, with robotic operations, under queue control, expected at the test facility in early 2015.

\section{Commissioning}
\label{sec:commissioning}
The commissioning of all four PlaneWave CDK-700 telescopes was performed from either the Caltech commissioning site ($34^\circ$ $08^\prime$ $10.0^{\prime\prime}$ N, $118^\circ$ $07^\prime$ $34.5^{\prime\prime}$ W; elevation $\approx 800$\,ft) or from the PlaneWave warehouse in Rancho Dominguez, CA ($33^\circ$ $52^\prime$ $14.1^{\prime\prime}$ N, $118^\circ$ $14^\prime$ $49.4^{\prime\prime}$ W; elevation $\approx 100$\,ft). The basic functionality of the telescopes including software control via the PlaneWave Interface software (PWI) has been validated through frequent use beginning 2013 April 12 when the first telescope was delivered to the Caltech campus, and extending through the writing of this publication. In the following sections we present the procedures and results for tests of telescope throughput and vignetting as well as fiber throughput and guiding.

\subsection{Telescope Throughput}
\label{sec:telthroughput}
The equation for the photoelectron detection rate is
\begin{multline}
S = \int_{\lambda_{\rm low}}^{\lambda_{\rm high}}A_{\rm eff}\tau(\lambda)f(\lambda)\eta(\lambda)F_\lambda(0) \times \\ 
10^{-0.4m}\exp\left[-\alpha(\lambda)X\right]\frac{\lambda}{hcg}{\rm d}\lambda
\label{eq:counts}
\end{multline}

where $S$ is the rate of detection of photoelectrons in ADU/s for a source with apparent magnitude $m$ relative to a zeropoint flux scale of $F_\lambda(0)$ at an airmass $X$. The quantity $A_{\rm eff}$ is the effective aperture---2998.33\,cm$^2$ for the CDK700 corresponding to 22\% central obscuration by the secondary, $f(\lambda)$ is the filter transmission function, $\eta(\lambda)$ is the camera quantum efficiency, and $g$ is the camera gain in units of $e^-$/ADU. The $\alpha$ factor is directly proportional to the atmospheric extinction with a proportionality constant of $0.4\ln(10)$, and we use $X = \sec(z)$ to estimate airmass which is sufficiently accurate for the range of zenith angles under consideration. We show explicitly the wavelength dependence of all quantities, and $h$ and $c$ are Planck's constant and the speed of light, respectively. The efficiency term, $\tau(\lambda)$, characterizes all sources of attenuation not accounted for by the other terms in Equation~\ref{eq:counts}, such as the reflectivity of the mirrors and the transmission of the lenses.

\begin{figure}
\includegraphics[width=\columnwidth]{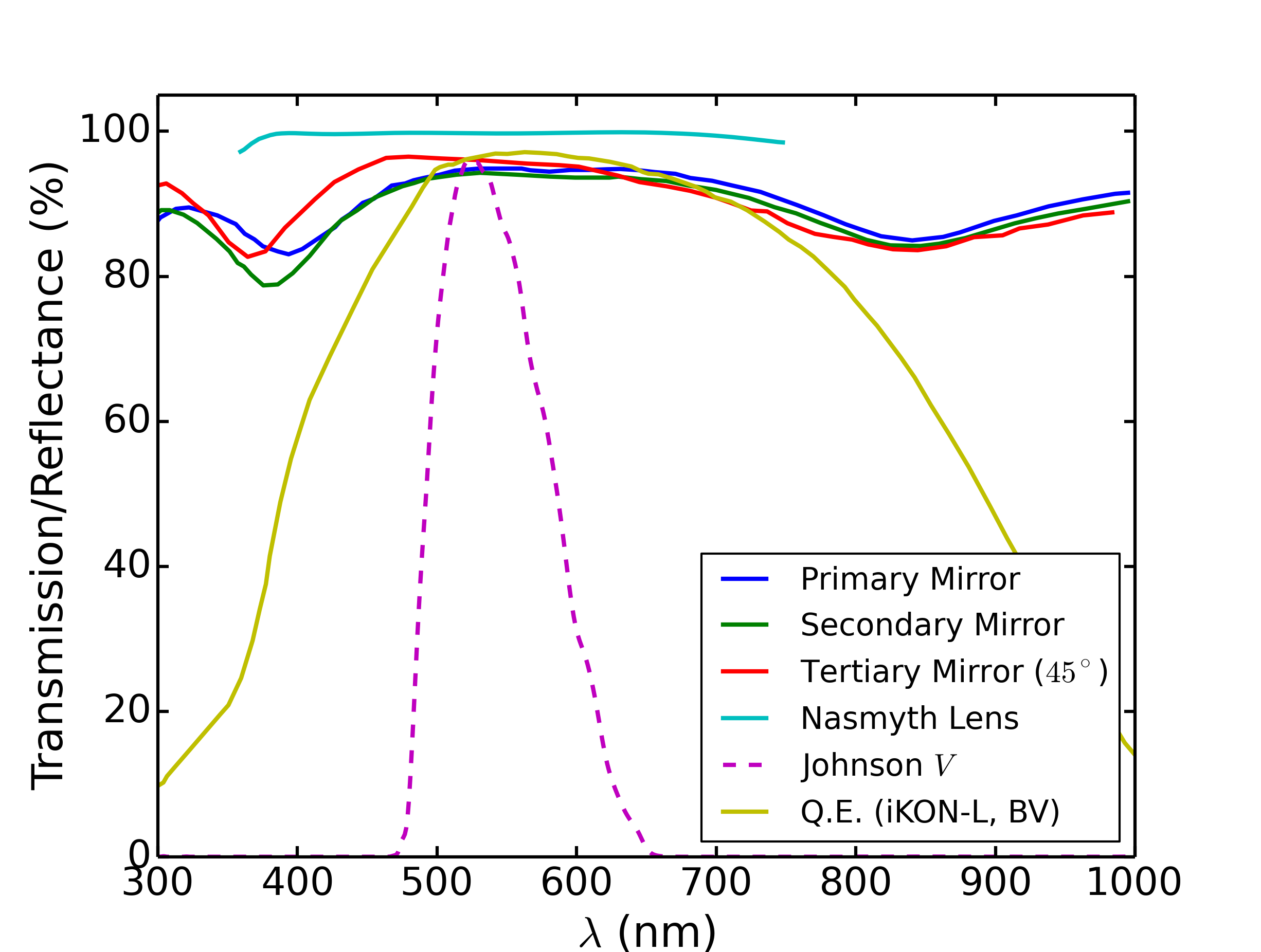}
\caption{\label{fig:trans}Transmission and reflectance of our optical elements as a function of wavelength. The reflectance of the tertiary mirror was measured at a $45^\circ$ incident angle, and the Nasmyth lens transmission is for a single lens surface. The quantum efficiency of the Andor camera (BV chip) is shown for reference, as is the Johnson $V$ bandpass which roughly corresponds to the spectral range of the MINERVA spectrograph.} 
\end{figure}

Camera gain was directly measured from flat field images and the theoretical transmission and reflectance of our optical elements were supplied by the manufacturers and are shown in Figure~\ref{fig:trans}. Multiplying these curves together (including the Nasmyth lens curve 4 times to account for the 4 surfaces on 2 lenses) and then integrating over the the $V$ filter band gives a theoretical upper limit to the telescope throughput of $\langle\tau\rangle_V(X=0) = 84$\%. 

To calibrate the throughput of our telescopes we observed bright ($V < 12$) standard stars\citep{Landolt1992,Landolt2013}. The standard stars were observed in sequences of between 3 and 10 images per pointing with integration times adjusted to give high counts within the linear regime of the CCD. The signal to noise achieved at each observation was typically a few hundred. We then cycled through a list of standard stars for a given night between 3 and 5 times producing measurements over a range of airmasses and at different azimuthal angles.

\begin{figure}
\includegraphics[width=\columnwidth]{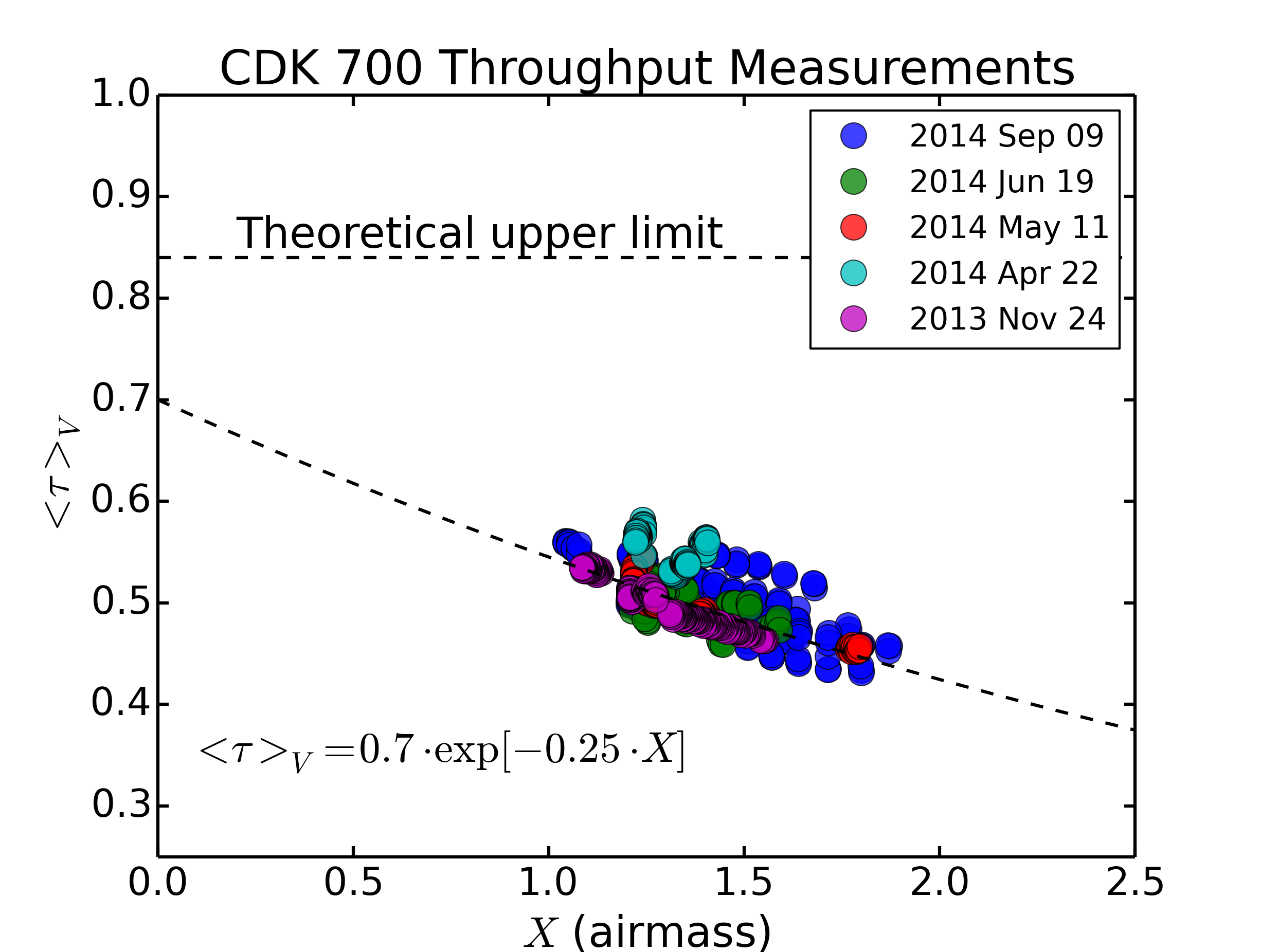}
\caption{\label{fig:throughput}Summary and fit of our throughput observations. The total throughput of the CDK-700 optics is estimated to be approximately $70\%$ in agreement with expectations.} 
\end{figure}

The total counts measured within the integration time of each observation were converted into a $V$-band averaged throughput for a calculated airmass by rearranging Equation~\ref{eq:counts}
\begin{equation}
\langle\tau\rangle_V = \frac{Shcg}{\int_{\lambda_{\rm low}}^{\lambda_{\rm high}} A_{\rm eff}f(\lambda)\eta(\lambda)F_\lambda(0)10^{-0.4m}\lambda {\rm d}\lambda}
\label{eq:throughput}
\end{equation}
We were able to collect data over airmasses between 1 and 2 with limited horizons at both the Caltech and Rancho Dominguez sites. Observations were spread as widely in azimuthal angle as possible on a given night and we repeated observations several times to average over a large variance in the individual measurements that we attribute to a highly variable aerosol content in the Los Angeles basin atmosphere. Figure~\ref{fig:throughput} shows the cumulative results of our throughput measurements using different telescopes and different cameras over the course of 5 separate nights. The overall best fit for the mean telescope throughput is 70\% with a high formal error of 30\% due to the covariance between the atmospheric extinction---which is poorly constrained by our data---and the throughput scale. However, the best fit value for the $V$ band atmospheric extinction, 0.27\,mags/airmass, is close to what we would expect and lends credence to the derived throughput of $\approx 70$\%. Fits from individual nights and sources using different telescopes and cameras also agree with this value within errors. 

\subsection{Vignetting}
\label{sec:vignetting}
\begin{figure}
\includegraphics[width=\columnwidth]{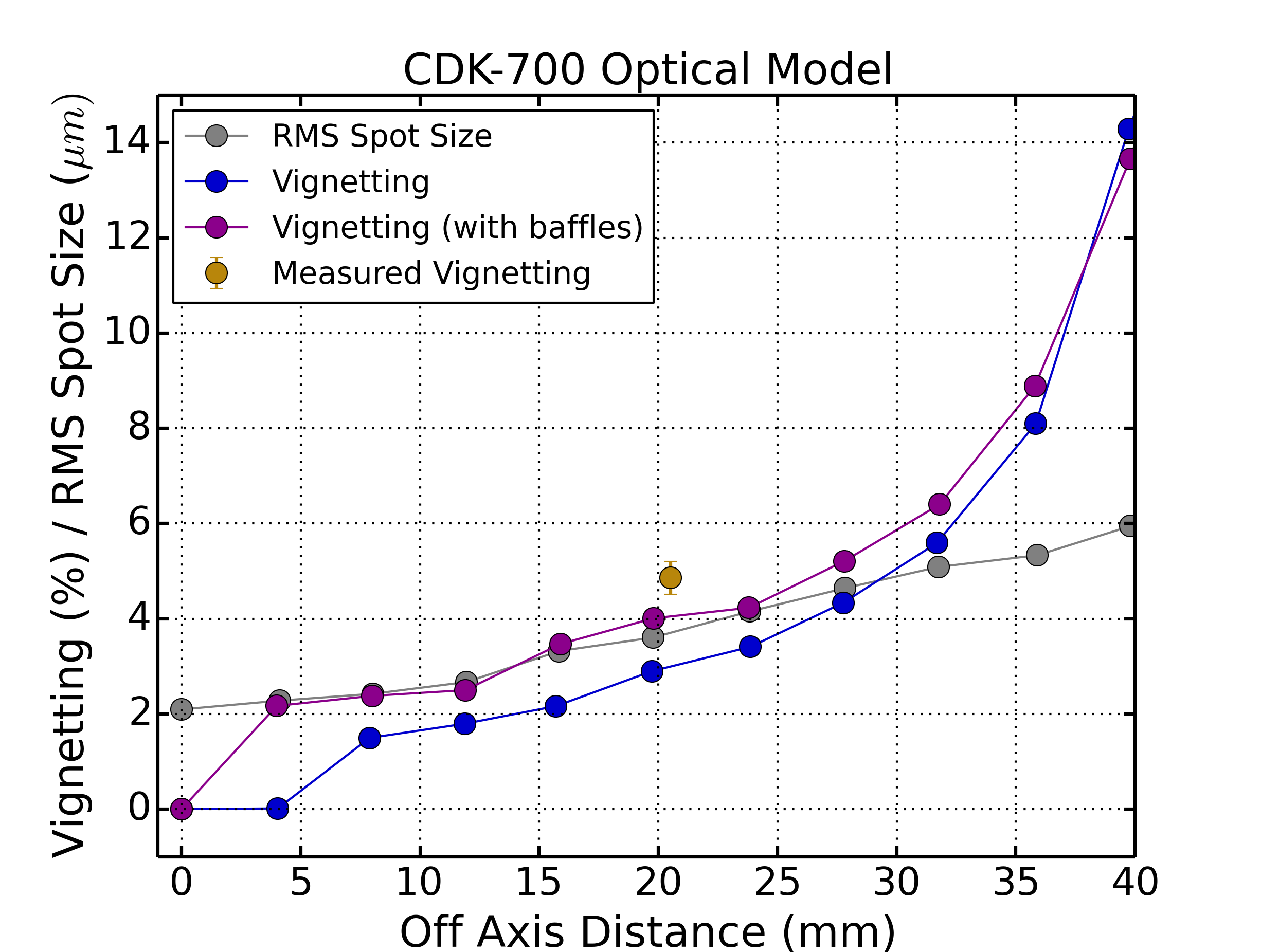}
\caption{\label{fig:vigmeas}Optical model of the CDK-700 provided by PlaneWave Inc. showing the expected level of vignetting in percent as a function of off axis distance ({\it blue dots and line}) and the RMS spot size in microns ({\it gray dots and line}). The on-sky measurement of Section~\ref{sec:vignetting} is shown in gold.} 
\end{figure}

To characterize the level of vignetting along our optical path we perform astronomical observations and compare with the optical model of the telescope. The optical model of the telescope is summarized in Figure~\ref{fig:vigmeas} where the vignetting percentage with and without mirror baffles and RMS spot size is plotted as a function of the distance from the optical axis in the focal plane. For the f/6.5 optics of the CDK-700, this corresponds to $45.3^{\prime\prime}$ per millimeter.

\begin{figure}
\includegraphics[width=\columnwidth]{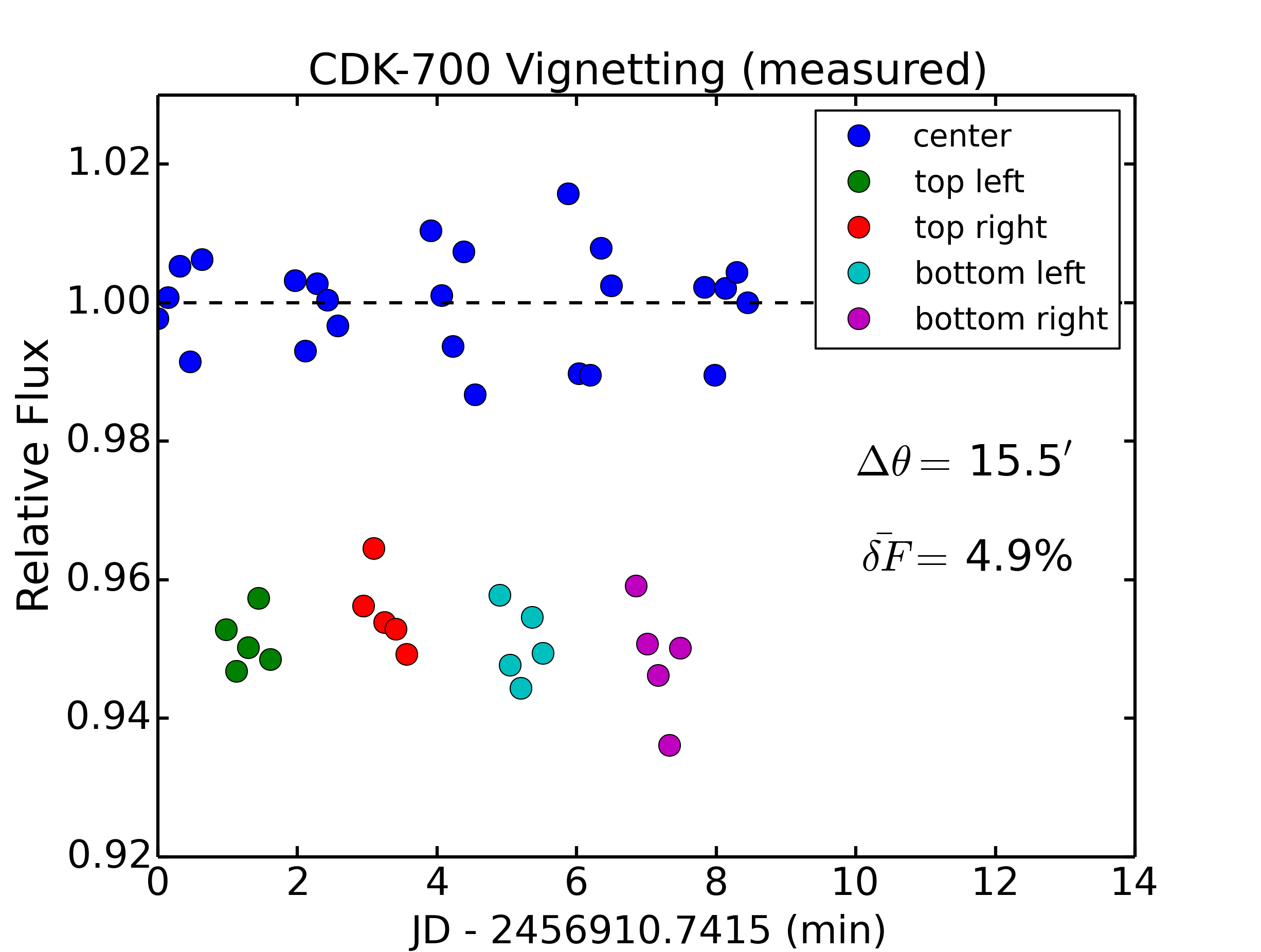}
\caption{\label{fig:vignetting}Results from our on sky vignetting test. A bright standard star was observed on 2014 September 9 alternately between the field center and the four corners of the CCD chip with telescope baffles in place. After correcting for zeropoint drift due to atmospheric changes we derive a $4.9\%\pm0.3\%$ relative vignetting at an off axis distance of $15.5^\prime$.} 
\end{figure}

The vignetting experiments used a bright standard star at high elevation, SA 111 773 ($V = 8.97$), observed alternately at the center of the CCD image and near to the four corners of the chip, $15.5^\prime$ off the field center. We use standard aperture photometry to derive the flux of the star in each of the positions, and then fit a polynomial function to the measured flux values of the standard star at the center of the chip to account for varying atmospheric conditions over the course of the observations. These variations were at the 1\% level. The flux values normalized to the polynomial fit reveal the relative flux decrement observed with the star at the corners of the CCD. The results from our observations using the fourth MINERVA telescope performed on 2014 September 9 are shown in Figure~\ref{fig:vignetting}. The average vignetting measured at the off axis positions is $4.9\%\pm0.3\%$. This value is approximatley 0.8\% above the expected level which is a statistically significant discrepancy given our measurement accuracy. However, this level of vignetting at the edge of our photometric field can be calibrated straightforwardly and is not expected to adversely affect our science goals.

\subsection{Pointing, Guiding and Source Acquisition}
\label{sec:pointing}
The pointing and guiding performance of the CDK-700 is dependent on a pointing model that converts astronomical coordinates into altitude and azimuth positions. The MINERVA commissioning site has limited sky access due to buildings and foliage preventing a full sky pointing model. Despite this limitation, we have found that the pointing and guiding of our telescopes typically achieve or exceed the specifications of Table~\ref{tab:telescope}. This level of performance surpasses our requirements to place our RV target stars within the $2^\prime$ field of view of our active guiding cameras.

The median seeing at the Mt. Hopkins Ridge site is about a factor of two better than our fiber diameters ($1.2^{\prime\prime}$ vs. $2.3^{\prime\prime}$), ensuring that minimal flux will be lost at the wings of the seeing disk given adequate guiding. The CDK-700 telescope open-loop tracking is accurate to a couple of arc seconds over the typical integration times of our primary program. Therefore, we have implemented a modified positional PID-type controller to correct for drifts and other inaccuracies. The controller input is the star position on the camera, and the output is altitude and azimuth offsets to the telescope mount. It typically sends corrections once every few seconds. The optical design of the telescope allows us to perform a one-time calibration of the camera field rotation if the telescope derotator is turned off, which is ideal for high-cadence observing. 

Once a target star arrives within the field of view of our guide cameras, the controller actively guides the telescope such that the star is placed on the calibrated pixel location of the fiber tip. The fiber tip location can be calibrated as frequently as needed to minimize losses due to offsets between the guiding center and the true center of the fiber. We have tested the stability of the fiber position on the guide camera and have found no measurable drifts or systematic offsets on day timescales. We have not measured the temporal stability over timescales from days to weeks. However, we do not anticipate that frequent calibrations will be needed as the optics are rigid and the fiber is strain relieved. Also, for these bright stars, reflections off the fiber cladding during observations can be detected and used as a secondary check for optical alignment. 

Currently, the controller operates near the optimal level, showing an RMS pointing precision of about $0.2^{\prime\prime}$, dominated by uncertainties in the measurement from seeing variations. Simulations of the required pointing accuracy indicate that coupling penalties below 5\% are incurred for a pointing accuracy of $0.2^{\prime\prime}$ RMS at any seeing from $0.5^{\prime\prime}$ to $2.5^{\prime\prime}$. It is thus unlikely that the control system is contributing to any major loss of throughput in the system. The most convincing evidence that the controller is not adding significant noise to the pointing system is provided by the amplitude spectral density of the pointing errors; the error is comparable or lower at all sensed frequencies when the telescope is guiding \citep[see][Figure 2]{Bottom2014}). Typical guide camera exposure times are 0.1\,s for stars from 4--6\,mag allowing sufficient sensitivity to successfully guide on the dimmest targets in our target list.

\begin{figure}
\includegraphics[width=\columnwidth]{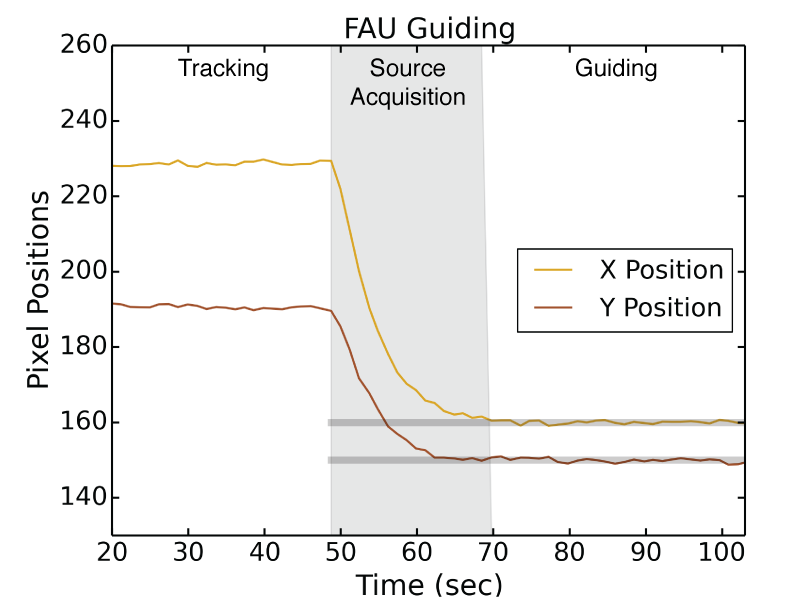}
\caption{\label{fig:guiding} Results from the FAU guiding test \citep{Bottom2014}, here showing a zoom in of the source acquistion. The active guiding system was initiated around 49 seconds, and the position of the star stabilized on the target pixel in approximately 20 seconds.} 
\end{figure}

Results from a guiding test are shown in Figure~\ref{fig:guiding}. For this test, first presented by \cite{Bottom2014}, the location of a bright star was tracked on the FAU guide camera up until about 49 seconds, after which active guiding was initiated. The time to acquire the source on the chosen pixel (here $x,y = 160,150$) took approximately 20 seconds. Although this controller has not been optimized to minimize the acquisition time, we use this result as the basis for total source acquisition time including telescope slew time.

\subsection{Fiber Coupling and Throughput}
\label{sec:fiberthroughput}
The theoretical throughput curve for our 50\,\micron\ octagonal fibers has been calculated as a function of astronomical seeing based on the vendor supplied transmission specifications for the pellicle, the fiber transmission, and the expected transmission calculated for the input and output reflectance \citep[see][Figure 3]{Bottom2014}. On-sky tests of the fiber coupling device performed from the MINERVA commissioning site on the Caltech campus in $\sim 2^{\prime\prime}$ seeing conditions show very good agreement with these expected values in consideration of fiber losses, reflection losses and coupling efficiencies. 

On-sky throughput observations performed on the Caltech commissioning site have confirmed the expected performance of the FAU design achieving 50\% measured efficiency (45\% throughput). Further tests will be needed at FLWO to validate the performance at more optimal conditions. The final version of this instrument will be deployed later this year, and will incorporate minor modifications such as a customized pellicle for slightly higher transmission (98\% vs. 92\%). These results suggest that the throughput of our fiber system will be roughly 70\% at FLWO where the median seeing is $1.2^{\prime\prime}$. With these results and those of Section~\ref{sec:telthroughput}, we expect to lose approximately 50\% of astronomical light from our telescope and fiber systems excluding losses from the atmosphere. 

\section{First Science Results}
\label{sec:science}
While the fair weather of the Los Angeles basin allows for routine commissioning operations, the Caltech site is a challenging place from which to obtain science grade astronomical data. Despite difficulties involving a highly variable atmosphere, significant obstruction, and copious stray light, we have been able to surpass our lower limit for photometric precision required by our secondary science objectives from this location. We also present new observations of WASP-52b, a transiting hot-Jupiter, as an end to end test of the MINERVA photometry pipeline.

\subsection{High-Precision Photometry of 16\,Cygnus}
\label{sec:phot}
The secondary science goal of MINERVA is to search the transit windows of known and newly discovered super-Earths detected by the radial velocity technique, including potential detections from the MINERVA target list. The transit of a 3\,\rearth\ planet around a Sun-like star ($0.8 \lesssim M_\star/M_\odot \lesssim 1.2$) produces a decrement of light on the order of one mmag. This is an approximate upper bound of what would be considered a ``super-Earth'' and therefore represents a lower limit to the precision that must be achieved with MINERVA for our secondary science program to be viable. The timescale for this precision is also important. A super-Earth in its respective Habitable Zone of a Sun-like star will transit with a duration of about 13\,hrs and have an ingress/egress time of about 20\,min. Planets closer to their host star are easier to detect, have higher transit probabilities and shorter transit durations. Therefore this level of precision should be attained on timescales less than $20$\,min.

On the evening of UT 2014 September 18 we observed one of the $\eta_\oplus$ targets that will be part of our RV survey, 16\,Cyg\,AB ($V = 5.95/6.20$, also HD\,186408/186427). The first MINERVA telescope was equipped with an Andor iKON-L camera, an SBIG ST-i guide camera, and a 7 slot filter wheel. The telescope was controlled through the PWI interface, while the camera and active guiding were controlled through Maxim DL\footnote{\url{http://www.cyanogen.com/maxim\_main.php}}. A series of flats were taken during twilight. There were some clouds in the East, but the area of sky where we were observing remained clear throughout our observations. 

\begin{figure}
\includegraphics[width=\columnwidth]{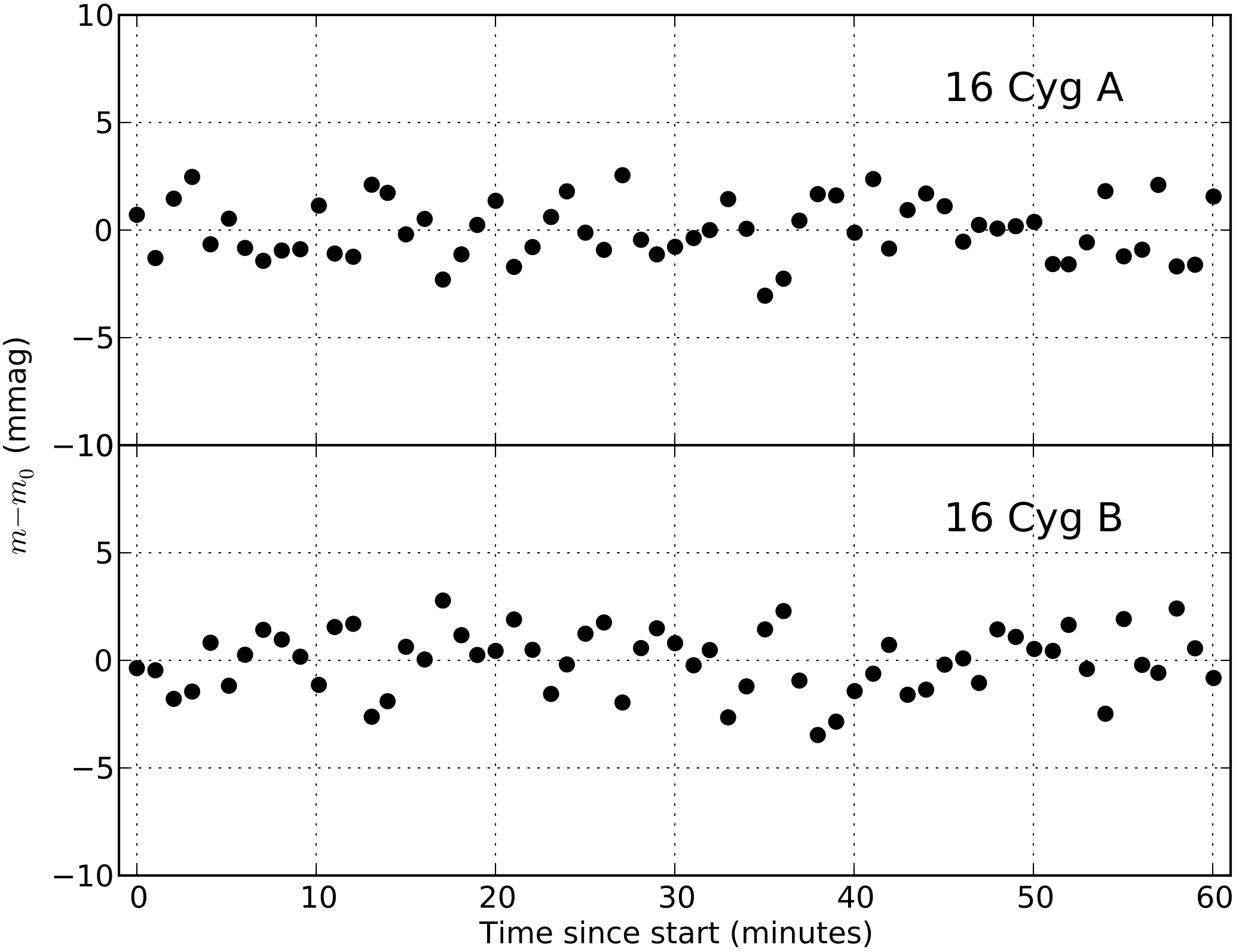}
\caption{\label{fig:16Cyg_lc}Detrended photometric time series of the 16 Cyg A and B observations performed on UT 2014 September 18 from the MINERVA commissioning site in Pasadena, CA. The individual seven second exposures have been binned into one minute intervals.} 
\end{figure} 
\begin{figure}
\includegraphics[width=\columnwidth]{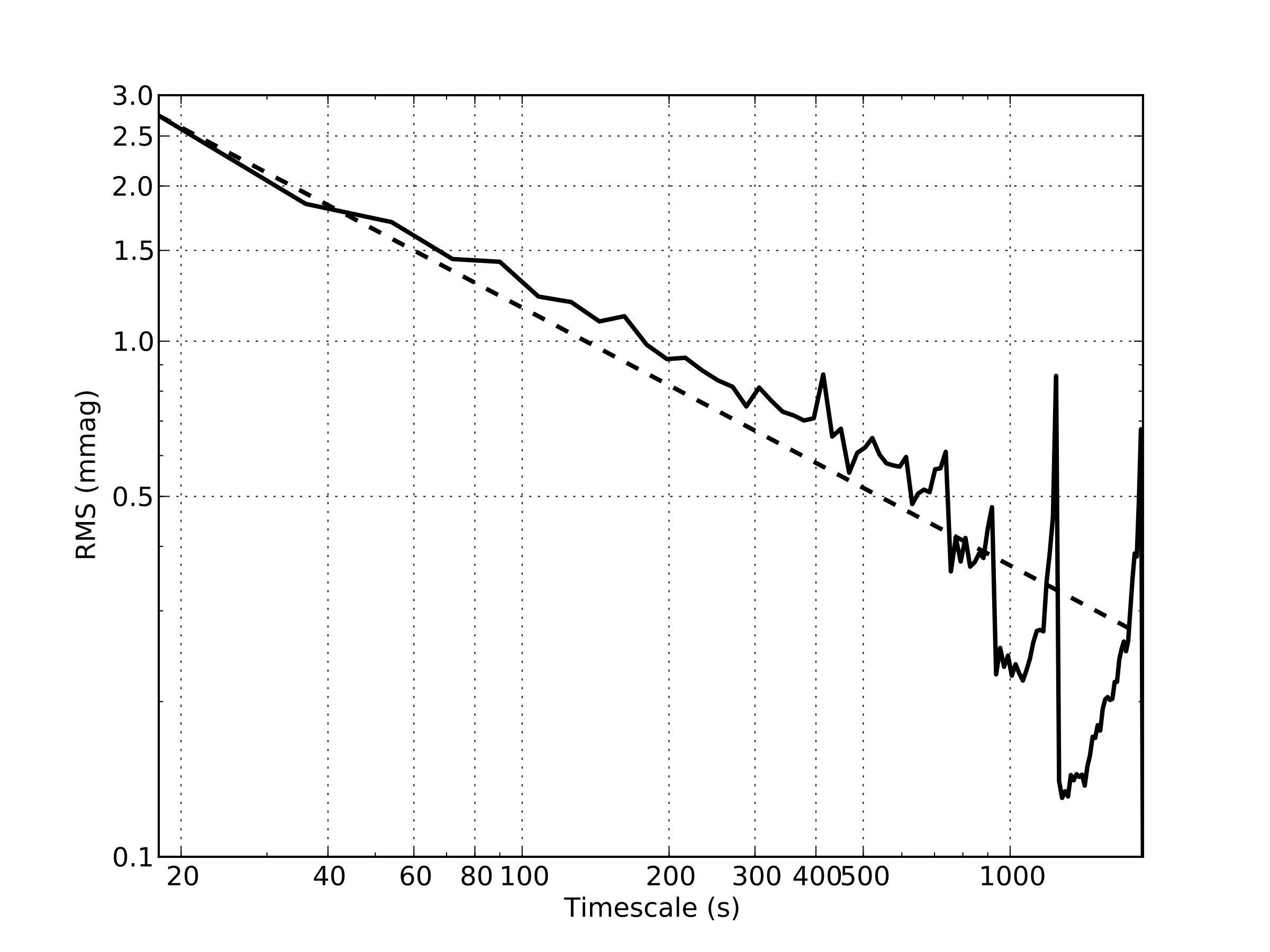}
\caption{\label{fig:16Cyg_allan}Allan Variance plot of the photometric time series of 16 Cygnus A performed on UT 2014 September 18 from the MINERVA commissioning site in Pasadena, CA. On the 17\,s duty cycle of our observations a 2.7\,mmag precision was achieved that bins down to below 1\,mmag on 3--5\,min timescales.} 
\end{figure}

Once on our target field, we aggressively defocused the camera while keeping our guide camera focused on a suitable guide star approximately $20^\prime$ off axis. The defocusing allows for the ample flux from our target to be spread over many more pixels mitigating the photometric error contributions from pixel-to-pixel variations and Poisson noise from the target itself. A second and important benefit from defocusing is to allow for long enough integration times to preclude shutter effects. To this second end, we also chose to observe in $z'$ band where the quantum efficiency of our camera is a factor of 2 below peak. A series of bias and dark frames were taken with the same setup following our observations.

Using 7\,s integrations and full frame readouts, we obtained 208 images of the 16 Cyg field spanning almost exactly an hour from UT 0352 to UT 0453. We actively guided throughout the course of these observations, but we did not achieve optimal guiding results. A drift of a few pixels was seen over the course of these observations and there was one episode of a fairly large (about 7 pixel or $4^{\prime\prime}$) guiding excursion that took place over approximately 1.5 minutes. The frames were bias and dark subtracted, and divided by our calibrated median twilight flats to correct for pixel-to-pixel variations. 

To extract the photometry from the calibrated science images we used the multi-aperture mode of AstroImageJ \citep{Collins2013}, which uses simple aperture photometry and sky-background subtraction. For all of the stars for which we measured a lightcurve, we used a constant aperture size of 30 pixels ($18.4^{\prime\prime}$) and a sky annulus with an inner radius of 90 pixels and an outer radius of 100 pixels. The rather large sky annulus was necessary so that the background annuli centered on each of the stars in 16 Cyg did not include the other member of the binary. On each science image we recentered the apertures on the stellar centroids using the center-of-light method\citep{Howell2006}. 

We used a set of five nearby comparison stars to remove systematics in the lightcurves of 16 Cyg A and B. The set of comparison stars included the corresponding other member of the 16 Cyg binary, which provided effectively all of the comparison information, as the next brightest comparison star was approximately 3.5 magnitudes fainter. To remove any lingering systematic trends in the data, we then performed a linear detrending against airmass. 

The detrended photometric time series of both 16 Cyg A and B are shown in Figure~\ref{fig:16Cyg_lc}, where the individual points have been binned into one minute intervals. The Allan Variance of 16 Cyg A is plotted in Figure~\ref{fig:16Cyg_allan}. The RMS of the unbinned photometry is 2.7\,mmag, while we achieved sub-mmag precision on about 3--5 minute time scales. The stability of the atmosphere on Mt. Hopkins is considerably better than in Pasadena. Therefore, these first results from our commissioning site support the prospect of routinely achieving sub-mmag photometric precision from FLWO.

\subsection{WASP-52b: New Transit Observations and Modeling}
\label{sec:transit}
WASP-52b is an inflated hot Jupiter with $M = 0.5$\,M$_{\rm J}$ and $R = 1.3$\,R$_{\rm J}$ in a slightly misaligned, 1.75\,d orbit \citep{Hebrard2013}. The transits of WASP-52b were first observed by the SuperWASP survey \citep{Pollacco2006} in 2008 and 2009, and the most recent observations in the literature are precision light curves obtained in September 2011 \citep{Hebrard2013}. The host star is reported to have a mass of 0.87\,\msun\ and a rotation period of 11.8 days suggesting a gyrochronological age of 0.4\,Gyr. The age, small orbital distance, and obliquity of this planet may have important implications for the mechanism by which it formed and for the formation of hot-Jupiter systems in general \citep{Valsecchi2014b}.

We observed WASP-52 on the evening of UT 2014 September 18 from the MINERVA commissioning site on the Caltech campus in Pasadena, CA. It has a $V$ magnitude of approximately 12, more than 200 times fainter than 16 Cygnus. To maximize the signal to noise ratio of our observations of WASP-52, we again aggressively defocused. 
The observations were performed in $r'$ band and we used an integration time of 120 seconds to build up high signal to noise on the defocused star image. Active offset guiding was used throughout the observations, but the guide star was faint and the target star drifted by $\sim 15$ pixels or $9^{\prime\prime}$. This drift was little more than half the size of our defocused star images. The target was tracked for 4 hours 19 minutes starting approximately 13 minutes before the start of ingress. 
\begin{figure}
\includegraphics[width=\columnwidth]{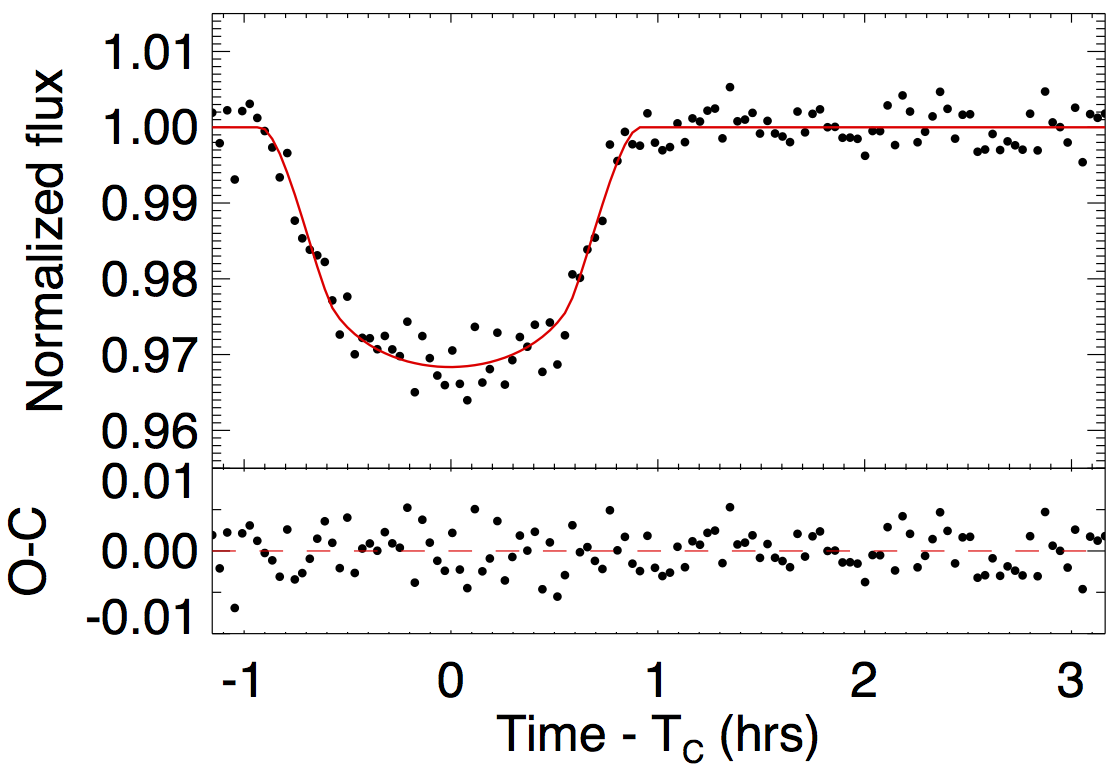}
\caption{\label{fig:WASP52_transit}Normalized and detrended light curve data for WASP-52 with best fit transit model overlaid. The transit center, $T_C$ is observed to be 6918.79085\,BJD$_{\rm TDB}$.} 
\end{figure} 

We used the same AstroImageJ \citep{Collins2013} reduction pipeline as in our 16 Cyg observations. We bias and dark subtracted our raw science images, before using a median twilight flat to remove image inhomogeneities. We then conducted simply aperture photometry with sky background subtraction on our calibrated images, and extracted lightcurves for WASP-52 and 12 other nearby comparison stars. For all of the stars which we extracted photometry we used a fixed 20 pixel aperture radius ($12.3^{\prime\prime}$) and a sky annulus with an inner radius of 30 pixels and an outer radius of 50 pixels. We recentered the apertures on the individual stellar centroids in each calibrated image using the center-of-light method \citep{Howell2006}. 

\begin{figure}
\includegraphics[width=\columnwidth]{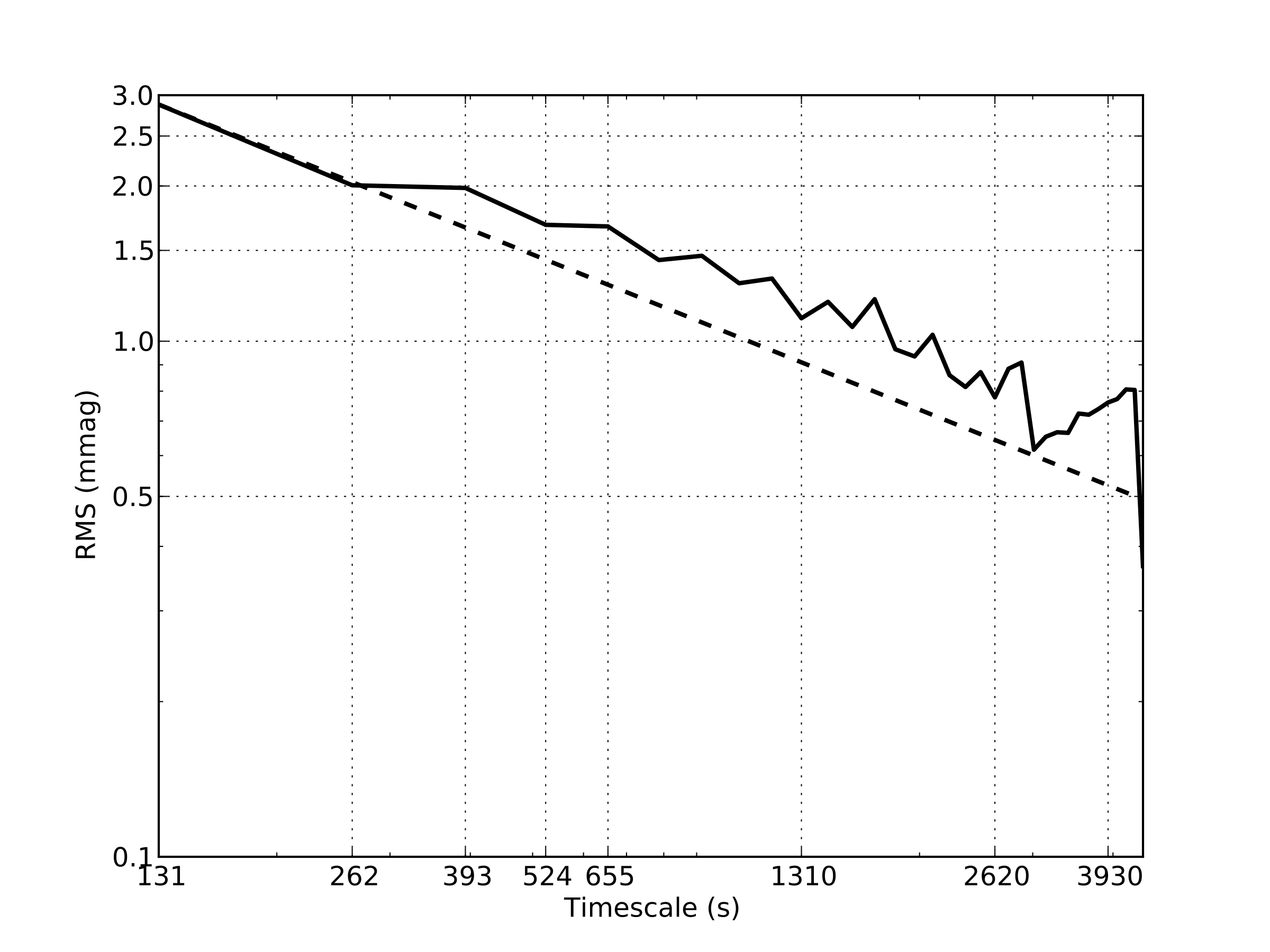}
\caption{\label{fig:WASP52_allan}Allan Variance plot of the photometric time series of WASP-52 performed on UT 2014 September 18 from the MINERVA commissioning site in Pasadena, CA. On the 131\,s duty cycle of our observations a 2.7\,mmag precision was achieved that bins down to 1\,mmag on 30\,min timescales.} 
\end{figure}

Figure~\ref{fig:WASP52_transit} shows the calibrated and detrended photometry of WASP-52. We achieved 3\,mmag precision on the 131\,s duty cycle of these observations which binned down to approximately 1\,mmag on 30 minute timescales. The Allan Variance for this photometric time series can be seen in Figure~\ref{fig:WASP52_allan}.

The midpoint Julian Date in Coordinated Universal Time (JD$_{\rm UTC}$) of each integration was recorded for each observation, which we convert to BJD$_{\rm TDB}$ \citep{Eastman2010}. The light curve of WASP-52 was then fit using EXOFAST \citep{Eastman2013}. We employ priors on the planet period, $P$; the stellar age, $Age$; metallicity, [Fe/H]; and effective temperature, $T_{\rm eff}$; all taken from \cite{Hebrard2013}. The transit parameters are tied to the Yonsei-Yale stellar models \citep{Demarque2004} through the stellar density as determined by the scaled semimajor axis, $a/R_\star$. This in turn informs a prior for a quadratic stellar limb darkening model parameterized by $u_1$ and $u_2$\citep{Claret2011}. The remaining free parameters of the fit are the baseline flux, $F_0$; transit time, $T_{\rm C}$; $\cos{i}$; $R_P/R_\star$; $\log{M_\star}$; an additional noise term added in quadrature to the transit errors, $\sigma_r^2$; and an error scaling for the uncertainties in the photometric time series, $TranScale$. While the differential photometry technique accounts for the majority of the airmass effects on our photometry measurements, additional drift is noticed that we suspect is due to scattered light effects. We therefore include an additional nuisance parameter in the fit that accounts for a linear drift with airmass.

The differential evolution MCMC sampler required 2080 burn-in steps and 13074 steps after the burn-in for adequate mixing \citep{Eastman2013}. Figure~\ref{fig:WASP52_transit} shows the best fit transit model overlaid on the detrended light curve data. The median parameter values and 1\,$\sigma$ errors are reported in Table~\ref{tab:WASP52}. 

Our parameter distributions are in broad agreement with those of \cite{Hebrard2013}. However, 1 to 1.5$\,\sigma$ discrepancies are seen between the scaled semimajor axis of the orbit, $a/R_\star$, and associated parameters. This discrepancy may arise from the difference in parametrization of the stellar limb darkening (Hebrard et al. use a 4 parameter limb darkening model, we use 2) or from the MCMC chains stepping in different parameters which introduces different priors.

The mid-transit time we derive for WASP-52b is $T_{\rm C} = 23 \pm 31$\,s advanced from the predicted mid-transit time, 643 transits from the literature ephemeris. This is consistent with the published ephemeris and extends the time baseline of WASP-52b transit photometry to approximately 6 years. The possibility of a third companion was mentioned by \cite{Hebrard2013} based on a potential acceleration of 30\,\ms\,yr$^{-1}$ seen in the radial velocity data. If real, this could be due to a massive giant planet at a few to several AU or a stellar companion with mass less than 0.8\,\msun\ further out \citep{Torres1999}. At those large orbital distances, the light travel time effect would dominate a transit timing variation (TTV) signal \citep{Borkovits2011}. Using the lack of TTVs in our data as a constraint, we can rule out the existence of brown dwarfs, $M \gtrsim 10$--15\,M$_{\rm Jup}$, out to approximately 6\,AU.

We also examine the photometry data from \cite{Hebrard2013} and find no significant variation of the transit depths in $r$ band for three epochs with full transit coverage. Independent fits to two epochs of data taken with the EulerCam on the 1.2\,m Euler-Swiss telescope in La Silla, Chile and one epoch of data from the 0.94\,m James Gregory Telescope in St. Andrews, Scotland agree with our transit depth to within 1\,$\sigma$. This suggests that WASP-52b may be a suitable target for follow-up observations of occultation events for the purpose of atmospheric studies \citep{Berta2012}. 

\newcommand{\bjdtdb}{\ensuremath{\rm {BJD_{TDB}}}}
\renewcommand{\feh}{\ensuremath{\left[{\rm Fe}/{\rm H}\right]}}
\renewcommand{\teff}{\ensuremath{T_{\rm eff}}}
\newcommand{\ecosw}{\ensuremath{e\cos{\omega_*}}}
\newcommand{\esinw}{\ensuremath{e\sin{\omega_*}}}
\renewcommand{\msun}{\ensuremath{\,M_\Sun}}
\renewcommand{\rsun}{\ensuremath{\,R_\Sun}}
\newcommand{\lsun}{\ensuremath{\,L_\Sun}}
\newcommand{\mj}{\ensuremath{\,M_{\rm J}}}
\newcommand{\rj}{\ensuremath{\,R_{\rm J}}}
\newcommand{\me}{\ensuremath{\,M_{\rm E}}}
\newcommand{\re}{\ensuremath{\,R_{\rm E}}}
\newcommand{\fave}{\langle F \rangle}
\newcommand{\fluxcgs}{10$^9$ erg s$^{-1}$ cm$^{-2}$}
\begin{deluxetable*}{lcc}
\tablecaption{Median values and 68\% confidence interval for WASP52b}
\tablehead{\colhead{~~~Parameter} & \colhead{Units} & \colhead{Value}}
\startdata
\sidehead{Stellar Parameters:}
                           ~~~$M_{*}$\dotfill &Mass (\msun)\dotfill & $0.842_{-0.045}^{+0.041}$\\
                         ~~~$R_{*}$\dotfill &Radius (\rsun)\dotfill & $0.732_{-0.042}^{+0.040}$\\
                     ~~~$L_{*}$\dotfill &Luminosity (\lsun)\dotfill & $0.305_{-0.051}^{+0.056}$\\
                         ~~~$\rho_*$\dotfill &Density (cgs)\dotfill & $3.03_{-0.33}^{+0.39}$\\
              ~~~$\log(g_*)$\dotfill &Surface gravity (cgs)\dotfill & $4.634_{-0.026}^{+0.028}$\\
                                ~~~$Age$\dotfill &Age (Gyr)\dotfill & $0.45_{-0.24}^{+0.27}$\\
              ~~~$\teff$\dotfill &Effective temperature (K)\dotfill & $5020\pm100$\\
                              ~~~$\feh$\dotfill &Metalicity\dotfill & $0.06\pm0.11$\\
\sidehead{Planetary Parameters:}
                              ~~~$P$\dotfill &Period (days)\dotfill & $1.7497798\pm0.0000012$\\
                       ~~~$a$\dotfill &Semi-major axis (AU)\dotfill & $0.02682_{-0.00049}^{+0.00043}$\\
                           ~~~$R_{P}$\dotfill &Radius (\rj)\dotfill & $1.166_{-0.088}^{+0.084}$\\
           ~~~$T_{eq}$\dotfill &Equilibrium Temperature (K)\dotfill & $1264_{-46}^{+45}$\\
               ~~~$\fave$\dotfill &Incident flux (\fluxcgs)\dotfill & $0.579_{-0.080}^{+0.086}$\\
\sidehead{RV Parameters:}
                   ~~~$T_A$\dotfill &Time of Ascending Node\dotfill & $2456918.35341\pm0.00036$\\
                  ~~~$T_D$\dotfill &Time of Descending Node\dotfill & $2456919.22830\pm0.00036$\\
\sidehead{Primary Transit Parameters:}
                ~~~$T_C$\dotfill &Time of transit (\bjdtdb)\dotfill & $2456918.79085\pm0.00036$\\
~~~$R_{P}/R_{*}$\dotfill &Radius of planet in stellar radii\dotfill & $0.1637_{-0.0039}^{+0.0035}$\\
     ~~~$a/R_{*}$\dotfill &Semi-major axis in stellar radii\dotfill & $7.88_{-0.29}^{+0.33}$\\
              ~~~$u_1$\dotfill &linear limb-darkening coeff\dotfill & $0.560\pm0.056$\\
           ~~~$u_2$\dotfill &quadratic limb-darkening coeff\dotfill & $0.174_{-0.054}^{+0.053}$\\
                      ~~~$i$\dotfill &Inclination (degrees)\dotfill & $86.52_{-0.63}^{+0.80}$\\
                           ~~~$b$\dotfill &Impact Parameter\dotfill & $0.479_{-0.095}^{+0.066}$\\
                         ~~~$\delta$\dotfill &Transit depth\dotfill & $0.0268_{-0.0013}^{+0.0012}$\\
                ~~~$T_{FWHM}$\dotfill &FWHM duration (days)\dotfill & $0.0620_{-0.0013}^{+0.0014}$\\
          ~~~$\tau$\dotfill &Ingress/egress duration (days)\dotfill & $0.0134\pm0.0014$\\
                 ~~~$T_{14}$\dotfill &Total duration (days)\dotfill & $0.0754_{-0.0012}^{+0.0013}$\\
      ~~~$P_{T}$\dotfill &A priori non-grazing transit prob\dotfill & $0.1061_{-0.0039}^{+0.0038}$\\
                ~~~$P_{T,G}$\dotfill &A priori transit prob\dotfill & $0.1477_{-0.0063}^{+0.0060}$\\
          ~~~$Depth$\dotfill &Flux decrement at mid transit\dotfill & $0.03155\pm0.00065$\\
              ~~~$d/R_*$\dotfill &Separation at mid transit\dotfill & $7.88_{-0.29}^{+0.33}$\\
     ~~~$\sigma_r^2$\dotfill &Variance of Transit red noise\dotfill & $-0.0000034_{-0.0000100}^{+0.0000083}$\\
         ~~~$Tran Scale$\dotfill &Scaling of Transit errors\dotfill & $1.14_{-0.66}^{+0.49}$\\
                   ~~~$\sigma_r$\dotfill &Transit red noise\dotfill & $0.0000_{-0.00}^{+0.0022}$\\
                            ~~~$F_0$\dotfill &Baseline flux\dotfill & $1.00003_{-0.00030}^{+0.00029}$\\
\sidehead{Secondary Eclipse Parameters:}
              ~~~$T_{S}$\dotfill &Time of eclipse (\bjdtdb)\dotfill & $2456919.66574\pm0.00036$
\enddata
\label{tab:WASP52}
\end{deluxetable*}

\section{Concluding Remarks}
\label{sec:conclusions}
The statistics from both RV and transit surveys of exoplanets have informed us that planets are very common throughout the Galaxy, and that the most common type of planet may be of a variety unrepresented in the Solar System---``super-Earths" with masses between that of the Earth and Neptune. The prevalence of these planets inferred from survey data imply that the stars in our local Solar Neighborhood should harbor many of these, with some fraction orbiting in their respective Habitable Zones. The radial velocity technique is the most promising method to date for detecting these planets. However, the detection of such small planets via the reflex motion of their host stars presents a significant challenge that requires cutting-edge instrumentation and a significant amount of observing time as the planet signals lie at or below the level of systematic noise generated from the stellar surface. 

MINERVA is an innovative facility designed to address these demands in a modular, cost-effective manner. By employing four small aperture telescopes from a commercial vendor we obtain a 1.4\,m effective aperture for a fraction of the cost of a single telescope. The small \'{e}tendue translates to a smaller spectrograph that is easier to stabilize requiring less infrastructure. Once the facility is complete, the array of telescopes will undertake an automated survey of a fixed target list over the course of several years that is expected to result in exoplanet detections of high scientific value.

We have presented the design and major components of MINERVA herein and the on-sky performance of our equipment up to and including the feeding of starlight into our fibers has been validated. Meanwhile, the procedures and data reduction pipeline for our secondary science goal of detecting transit events around nearby stars have been demonstrated with obseravtions from our test facility. There are, however, several more milestones the project will need to reach before science operations commence.

\subsection{Future Prospects: MINERVA-South and MINERVA-Red}
\label{sec:future}
In addition to the primary survey presented in this article, the modularity of the MINERVA design offers several opportunities for expansion that are already being pursued. The abundant yield of transiting exoplanets around bright stars in both the Northern and Southern hemispheres expected from the \kepler\ K2 ecliptic mission \citep{Howell2014} and the upcoming \textit{Transiting Exoplanet Survey Satellite} \citep[TESS;][]{Ricker2014} will require intensive RV follow-up. Planets with periods up to $\sim 100$\,d will be detected within the continuous viewing zone of TESS near the ecliptic poles that will require long term RV monitoring. As has been made clear from the \kepler\ prime mission, the most interesting planets will require a substantial amount of dedicated telescope time \citep{Howard2013,Pepe2013,Marcy2014}. Hence, there are significant opportunities for a MINERVA-like facility in the Southern hemisphere. MINERVA-South will take advantage of the same economies as described in Section 2, with the added advantage that our team's investment in software/hardware development and operational expertise will easily translate to the Southern facility. We expect MINERVA-South to be of essentially the same design as MINERVA, sited in Australia or Chile, and operational by 2018 to capitalize on the coming flood of planet candidates from K2 and TESS.

In the near term, we are expanding the reach of the MINERVA project to include nearby M stars with a second instrument specifically designed for gathering precise radial velocities of the closest low-mass stars to the Sun. Recent results from \kepler\ and ground based surveys indicate that compact systems of planets in orbit around low-mass stars may be extremely common \citep{Swift2013, Dressing2013,bonfils13}. Statistically speaking, we expect some of the closest stars to the Sun to host systems of planets. However, these small, cool stars are often too faint to observe at the optical wavelengths where most precision RV instruments operate, including the MINERVA spectrograph. MINERVA-Red, a parallel effort to the main MINERVA survey, will specifically target a small sample of nearby mid- to late-M stars. 

The MINERVA-Red instrument is a fiber-fed echelle spectrograph housed in a vacuum chamber and optimized to cover the 800 to 900\,nm spectral region at resolution of $R\approx55,000$. The instrument is designed around single-mode fiber input, which allows the instrument to be very compact and stable, and also eliminates modal noise as a source of RV error. While single-mode fiber reduces the possible coupling efficiency of starlight into the fiber, special attention has been paid to maximizing the optical throughput of the rest of the instrument. To that end, the spectrograph will operate with a pair of PlaneWave CDK-700 telescopes that have gold mirror coatings, significantly enhancing the telescope efficiency in the wavelength range of interest. These will be in addition to the 4 CDK-700s that will be used for the primary survey. The instrument optics are all optimized for this relatively narrow spectral range, and a Deep Depletion detector will ensure high Quantum Efficiency and low fringing. The MINERVA-Red instrument is currently under construction, and is slated to be deployed with the first of its two telescopes at the Mt. Hopkins site by mid-year 2015. 

\acknowledgments 
This work was partially supported by funding from the Center for Exoplanets and Habitable Worlds.  The Center for Exoplanets and Habitable Worlds is supported by the Pennsylvania State University, the Eberly College of Science, and the Pennsylvania Space Grant Consortium. MINERVA hardware has been partially funded by the Australian Research Council's \textit{Linkage, Infrastructure, Equipment and Facilities} funding scheme (project LE140100050). We are grateful to the Mt. Cuba Astronomical Foundation and the David and Lucile Packard Foundation for their generous funding of MINERVA hardware and personnel. CHB is supported by a NASA Nancy Grace Roman Technology Fellowship.

{\it Facilities:} \facility{MINERVA}

\clearpage

\end{document}